\documentclass[number,preprint,3p]{elsarticle}

\usepackage{color}
\usepackage{graphicx}
\usepackage{subcaption}
\usepackage{algorithm}
\usepackage{bm}
\usepackage[colorlinks]{hyperref}
\usepackage{amssymb}
\usepackage{amsthm}
\usepackage{amsmath}
\usepackage{amssymb}
\usepackage{mathtools}
\usepackage{dsfont}
\usepackage{booktabs}
\usepackage{url}
\usepackage{scrextend}
\usepackage{epstopdf}
\usepackage{float}
\usepackage{tcolorbox}
\usepackage{tablefootnote}
\usepackage[perpage]{footmisc}
\usepackage{bbm}

\newcommand{\defi}{\coloneqq}

\newcommand{\tr}{^\top}


\biboptions{square}
\journal{arXiv}

\begin{document}

\begin{frontmatter}

\title{The dynamics of entropy in the COVID-19 outbreaks}
  \author[1]{Ziqi Wang}
 \author[2,3]{Marco Broccardo}
 \author[4,5]{Arnaud Mignan}
 \author[6,4]{Didier Sornette}
 \address[1]{Earthquake Engineering Research and Test Center, Guangzhou University, China}
  \address[2]{Department of Civil Engineering and Industrial Design, University of Liverpool, United Kingdom}
  \address[3]{Department of Civil, Environmental and Mechanical Engineering, University of Trento, Italy} 
  \address[4]{Institute of Risk Analysis, Prediction and Management, Southern University of Science and Technology, China}
  \address[5]{Department of Earth and Space Sciences, Southern University of Science and Technology, China}
  \address[6]{Chair of Entrepreneurial Risks, Department of Management, Technology, and Economics, ETH Z{\"u}rich, Switzerland}

 \footnotetext[1]{Corresponding author: Ziqi Wang; \href{mailto:ziqidwang@yahoo.com}{ziqidwang@yahoo.com}}
 \footnotetext[2]{Corresponding author: Marco Broccardo; \href{mailto:bromarco@ethz.ch}{bromarco@ethz.ch}}

\begin{abstract}
With the unfolding of the COVID-19 pandemic, mathematical modeling of epidemics has been perceived and used as a central element in understanding, predicting, and governing the pandemic event. However, soon it became clear that long term predictions were extremely challenging to address. In addition, it is still unclear which metric shall be used for a global description of the evolution of the outbreaks. Yet a robust modeling of pandemic dynamics and a consistent choice of the transmission metric is crucial for an in-depth understanding of the macroscopic phenomenology and better-informed mitigation strategies.
In this study, we propose a Markovian stochastic framework designed to describe the evolution of entropy during the COVID-19 pandemic and the instantaneous reproductive ratio. We then introduce and use entropy-based metrics of global transmission to measure the impact and temporal evolution of a pandemic event. In the formulation of the model, the temporal evolution of the outbreak is modeled by the master equation of a nonlinear Markov process for a statistically averaged individual, leading to a clear physical interpretation. The time-dependent parameters are formulated by adaptive basis functions, leading to a parsimonious representation. We also provide a full Bayesian inversion scheme for calibration and a coherent strategy to address data unreliability. The time evolution of the entropy rate, the absolute change in the system entropy, and the instantaneous reproductive ratio are natural and transparent outputs of this framework. The framework has the appealing property of being applicable to any compartmental epidemic model.
As an illustration, we apply the proposed approach to a simple modification of the Susceptible-Exposed-Infected-Removed (SEIR) model. Applying the model to the Hubei region,   South Korean, Italian, Spanish, German and French COVID-19 data-sets we discover significant difference in the absolute change of entropy but highly regular trends for both the entropy evolution and the instantaneous reproductive ratio.



\end{abstract}


\end{frontmatter}

\renewcommand{\thefootnote}{\fnsymbol{footnote}}

\section{Introduction}

\noindent Coronaviruses are one of the most significant threat to human society \cite{ClinicalFeatCov}\cite{Clinicalfeatureschildren}\cite{Clinicalfeaturespregnant}\cite{PerceptionsBehaviouralResponses}\cite{severitycoronavirusdisease}\cite{HUANG2020497}. Limited to short outbreaks in the recent past \cite{Firstknownpersontoperson}\cite{Firstwave}\cite{early-stageimportationrisk}, their pandemic-level potential was well known \cite{WHO}\cite{securitycapacities}, yet most countries proved unprepared to cope with the so-called coronavirus infectious disease of 2019 (COVID-19). Revealed in the Hubei province, China, the novel coronavirus has spread all over the world. China responded with massive containment measures starting at the end of January 2020, which limited further contamination on the mainland \cite{effectofcontrolstrategies}\cite{Firstwave}. In Europe, most individual states have responded with similar containment measures. However, there has been a lack of common European action. Strict or soft containment measures have been applied with different timeframes and specialized to individual health and socio-cultural systems,  showing very different pandemic evolutions. At the time of writing, the main episode of COVID-19 is (in general) under control in China, South Korea, and continental Europe \cite{WHO}, despite the possibility of multiple waves. On the contrary, North and South America are still in the middle of the pandemic, and a clear picture of the evolution of events is not possible yet. 

The amount of data available allows various modeling techniques to be tested more robustly than in previous epidemics. However, no model (from the physics-based to the purely data-driven) has been or is able to predict the long-term evolution of the pandemic accurately (conversely, short-term predictions are possible with some degree of accuracy \cite{ICLPredict}\cite{wu2020generalized}). There are several reasons behind this long-term unpredictability; an incomplete list includes the partial understanding of the phenomenon, the (many, or even infinite) missing variables, the high sensitivity of the model to parameters, the incomplete/inaccurate data acquisition scheme, and the lack of uniform measurement methods. However, a profound reason that makes any long-term prediction difficult is the presence of endogenous variables (a well known problem in social sciences \cite{Favero2020}). The endogenous variables may involve local policies, socio-cultural aspects, human behaviors, communication strategies, and they are typically difficult to model and measure. Since an epidemic evolves as a result of the interplay between the ``natural evolution'' of the disease and society/human interventions, a robust and generalizable  microscopic model with complete characterizations of endogenous variables is challenging to build. Given this, in this study, we use macroscopic phenomenology based modeling to gain insight into the epidemic dynamics\footnote{This approach follows the famous sentence of Richard Hamming ``The purpose of (scientific) computing is insight, not numbers."}. Therefore, here, the goal is not to provide long term numerical predictions, although the proposed modelling technique can be used for extrapolation.

Among various modelling options, Susceptible-Infected-Removed (SIR) types of compartmental models have gained wide popularity due to their simplicity and straightforwardness in interpreting the macroscopic phenomenology. A significant amount of SIR-type model based studies have already been carried out to investigate the transmission properties of the COVID-19, and an incomplete list includes \cite{Firstwave}\cite{effectofcontrolstrategies}\cite{PNASPandemic}\cite{earlyestimationofepidemiological}\cite{Phaseadjustedestimation}\cite{LancetNowcasting}\cite{ModifiedSEIRandAI}. The spectrum of complexity of these models is broad. They can range from a minimum number of compartments (which offers a better generalization) to a large number of compartments (which offers a better local description). They can be deterministic (i.e., counting the deterministic number of individuals for each compartment) or stochastic (i.e., defining a joint probability measure of the number of individuals for each compartment). They can have different data acquisition schemes (from a simple frequentist analysis of the single parameters to a complete Bayesian inversion scheme).  Finally, they can simply macroscopically describe the pandemic evolution of a given location (top-down approach), or include a spatial topological description (including mobility) and/or a different degree of spreading among individuals by including adjacent matrices  (bottom-up approach).

Given a data set, these models can be calibrated and offer new insights into the evolution of the pandemic. For example, they can shed light on how the pandemic developed by measuring the reproductive ratio  (constant or time-varying) and finally estimating the effectiveness of containment measures. This metric also allows for a comparison between different regions but does not provide a quantitative measure of the impact of the spread. On the other hand, the evolution of the number of infected and deaths provides a means of direct impact; however, they lack objectivity as they are strongly influenced by the different populations of the regions, the measurement strategies, and the unreliability of the data. Furthermore, they are not global metrics as they do not provide an objective and robust way to unify them into a single (scalar) measure.  Therefore, there is a research gap on how to provide a macroscopic model-metric pair to compare different regions' performance and get new insights into various outbreaks.

This study aims to fill this gap by proposing a macroscopic stochastic model equipped with a global transmission metric based on entropy. In this context, the entropy evolution of the process is a metric that describes the degree of disorder (i.e., of impact) of an epidemic. This metric allows for an objective comparison between regions and provides a global measure of both the evolution and the impact of COVID-19 outbreaks.  In particular, we propose a compartmental stochastic model that has the following characteristics. i) \textit{Stochastic}: the model describes a statistically averaged individual by a nonlinear Markov process with compartmental epidemic states. ii) \textit{Time-dependent}: the model parameters are decomposed onto generic basis functions (of time). iii) \textit{Parsimonious}: instead of conventional orthogonal basis functions (e.g., orthogonal polynomials, Fourier/wavelet series) the adaptive basis functions are adopted to achieve a representation with minimum number of basis functions. iv) \textit{Bayesian}: the time-dependent parameters are assumed to be random and are calibrated by full Bayesian inversion. Furthermore, we equip the model with a metric based on entropy which has the following characteristics. i) \textit{Meaningful}: the metric provides a physical and transparent measure of the COVID-19 impact in a given region; moreover, it is by definition the time integral of the entropy rate, which represents the temporal evolution of the epidemic. ii) \textit{Global}: the metric provides a global and average description of the pandemic event. iii) \textit{Consistent}: the metric is not influenced by the number of individuals, and can be used objectively to compare different regions. iv)\textit{Robust} the metric is associated with an error that is a direct output of the Bayesian inversion scheme used to calibrate the stochastic model. Finally, to have a reliable description of the events, we provide robust strategies to fill in missing information and to correct the numerous inconsistencies on the current data sets.

The paper is organized as follows. First, we develop general concepts of the proposed epidemic model, including governing equation, time-dependent parametrization and Bayesian model calibration (Section~\ref{Framework}). Second, we introduce the entropy-based metric in Section~\ref{Sec:Entropy}. Third, we apply the proposed approach to formulate a SEIR compartmental model for modelling the temporal evolution of COVID-19 (Section \ref{Application}). Next, we apply the proposed approach to real data-sets to the following regions: Hubei (China), South Korea, Italy, Spain, Germany and France (Section \ref{NumericalTest}). Finally, we conclude the study by identifying the limitations, conclusions and future research directions.

\section{The stochastic epidemic model}\label{Framework}
\noindent In the literature, the term ``stochastic compartmental model'' can refer to different formulations (see, e.g., \cite{ModelingInfectiousDiseases} for a review) with distinct underlying assumptions on the source of uncertainty. For instance, the noise-driven stochastic model is formulated by: i) introducing additive noise process into the deterministic compartmental model; ii) translating the noise into diffusion of probability distribution; iii) obtaining an equation of probability distribution (e.g., Fokker-Plank equation). Clearly, in the noise-driven model the source of uncertainty is the additive noise. An alternative and more popular stochastic formulation is the event-driven model, which can be summarized as a direct \textit{stochastic simulation} of the deterministic model. Specifically, in the event-driven model the deterministic rate matrix is used to define the transition probability of event $X_m(t)\xrightarrow{t+\Delta t}X_m(t)\pm \Delta X$, where $X_m(t)$ denotes the population in a compartment, and $\Delta X$ the intra-state increment. With the transition probability, a direct stochastic simulation (via e.g., Gillespie’s Direct Method \cite{GILLESPIE1976403}\cite{GillespieMethod}) would yield a random scenario of the epidemic. The proposed model can also be classified as an event-driven approach in the sense that the source of uncertainty is also the aleatory variability of transitions between epidemic states. However, instead of a stochastic simulation without a governing equation of probability distribution, the proposed model strictly follows an equation of probability distribution which describes a nonlinear Markov process. Consequently, the proposed model possesses clear physical interpretations within the mathematical framework of nonlinear Markov process theory.

Compartmental models with time-varying parameters have been widely studied in the literature \cite{Ebola}\cite{StochasticTimeVaryingSEIR}\\\cite{ExtractTimeDependent}\cite{TimeDependentSEIR}\cite{TimeDependentSIR}. A fundamental question to be addressed in time-dependent models is the trade-off between over-fitting and under-fitting, or equivalently, model bias versus model variance. In the extreme scenario, a pointwise kernel based parametrization may lead to an almost exact calibration on epidemic observations, yet the explanatory/extrapolation capability would be minimized, and the model variance would be maximized. In an over-parameterized model, the non-local trend/structure, which is crucial to characterize/understand the epidemic dynamics, can hardly be identified. In this study, we attempt to discover non-local structures from the epidemic dataset using a parsimonious formulation with adaptive basis functions. 
Moreover, since the model calibration is formulated in a Bayesian framework, likelihood-based model selection, e.g., using the Bayesian information criterion (BIC) \cite{BIC}, can be conveniently applied to specify the number of adaptive basis functions. 

\subsection{The original deterministic model}
\noindent Consider a generic compartmental epidemic model with a fixed\footnote{A fixed total population indicates a closed system, i.e., i) the vital dynamics (natural birth/death) is neglected, assuming that the course of the epidemic is relatively short; ii) the immigration and emigration are neglected, assuming that $N$ is sufficiently large and the course of immigration/emigration is relatively slow.} total population $N$ and a classification of the population into $M$ compartments $\bm X=[X_1,...,X_M]\tr$. The compartmental epidemic model describes the temporal evolution of the state vector $\bm X$, where every component of $\bm X$ is by definition nonnegative and $\bm X$ is subjected to the conservation law $\|\bm X\|_1=N$.

For an infinitesimal incremental $\Delta t$, we study the following master equation of state vector.
\begin{equation}\label{MasterEq}
\bm X(t+\Delta t)=(\bm I+\bm H(\bm X(t),t)\Delta t)\bm X(t)\,,\
\end{equation}
where $\bm I$ is the identity matrix and $\bm H(\bm X(t),t)$ is a problem-specific rate matrix (infinitesimal \textit{propagator}). Eq.\eqref{MasterEq} is equipped with the assumption that the evolution of $\bm X(t)$ is smooth\footnote{Here we have to assume the population is a real number rather than an integer, later in the probabilistic reformulation this assumption can be relaxed.}, i.e., without jumps. Setting $\Delta t\to0$, Eq.\eqref{MasterEq} leads to
\begin{equation}\label{MasterEqDiff}
\frac{d\bm X(t)}{dt}=\bm H(\bm X(t),t)\bm X(t)\,.\
\end{equation}
Similar to mechanics, the rate matrix $\bm H(\bm X(t),t)$ governs the dynamics of $\bm X(t)$. To preserve the conservation law $\|\bm X(t)\|_1=N$, we must have $\bm H(\bm X(t),t)\tr\bm 1=\bm 0$, where $\bm 1$ is a vector of ones and $\bm0$ the null vector.

Particularly, if $\bm X(t)$ eventually attains a stationary state $\bm X^*$ defined as
\begin{equation}\label{StationX}
\bm X^*\defi\lim_{t\to+\infty}\bm X(t)\,,\
\end{equation}
and define $\bm H^*$ as
\begin{equation}\label{StationR}
\bm H^*\defi\lim_{t\to+\infty}\bm H(\bm X(t),t)\,.\
\end{equation}
We obtain the stationarity condition
\begin{equation}\label{StationLaw}
\bm H^*\bm X^*=\bm0\,,\
\end{equation}
where $\bm0$ is a column vector of zeros.

\subsection{Probabilistic reformulation}
\noindent Given a deterministic $\bm H(\bm X(t),t)$, Eq.\eqref{MasterEqDiff} describes a deterministic trajectory of $\bm X(t)$. Since variabilities inevitably exist in the specification of $\bm H(\bm X(t),t)$ or/and the initial condition, the solution $\bm X(t)$ becomes a multivariate stochastic process. However, the aforementioned ``randomization'' is regarded as epistemic with respect to the model Eq.\eqref{MasterEqDiff}\footnote{The classification of epistemic or aleatory uncertainties should always be accompanied by the specification of a model universe. Here if we assume Eq.\eqref{MasterEqDiff} is the correct underlying (deterministic) law, the uncertainties of $\bm H(\bm X(t),t)$ is epistemic (with respect to Eq.\eqref{MasterEqDiff}) because if the form of $\bm H(\bm X(t),t)$ is fixed the law of Eq.\eqref{MasterEqDiff} is deterministic.}. This section focuses on a more fundamental (aleatory) probabilistic reformulation of Eq.\eqref{MasterEqDiff}.

Adopting a \textit{frequentist} point of view on probability, consider a normalization of $\bm X(t)$ by
\begin{equation}\label{Normalize}
\bm P(t)\defi\lim_{N\to+\infty}\frac{\bm X(t,N)}{\|\bm X(t,N)\|_1}=\lim_{N\to+\infty}\frac{\bm X(t,N)}{N}\,,\
\end{equation}
where $X(t,N)$ is used to highlight that the compartmental population depends on $N$, and $\bm P(t)$ can be interpreted as the marginal probability distribution of a discrete state continuous time stochastic process. Observe that despite $\bm P(t)$ being equivalent to proportions in a deterministic model, the probabilistic individualistic interpretation leads to a fully stochastic dynamic interpretation of the problem. The underlying state associated with $\bm P(t)$ is an epidemic state of a statistically averaged \textit{individual}. Analogous to Eq.\eqref{MasterEq} and Eq.\eqref{MasterEqDiff}, we obtain
\begin{equation}\label{PMasterEq}
\bm P(t+\Delta t)=(\bm I+\bm Q(\bm P(t),t)\Delta t)\bm P(t)\,,\
\end{equation}
and
\begin{equation}\label{PMasterEqDiff}
\frac{d\bm P(t)}{dt}=\bm Q(\bm P(t),t)\bm P(t)\,,\
\end{equation}
where $\bm Q(\bm P(t),t)$ is a rate matrix analogous to $\bm H(\bm X(t),t)$ in the deterministic model. Since $\bm Q(\bm P(t),t)$ explicitly depends on $\bm P(t)$, Eq.\eqref{PMasterEqDiff} describes a \textit{nonlinear Markov process} \cite{NonlinearMarkov}. The conservation of probability is guaranteed by $\bm Q(t)\tr\bm 1=\bm 0$.

Eq.\eqref{PMasterEq} provides a straightforward strategy for sampling random realizations of the process. In particular, for a fixed initial condition $\bm P(t_0)$, the solution $\bm P(t)$ is deterministic, and $\bm Q(\bm P(t),t)$ can be regarded as $\bm Q(t)$ with $\bm P(t)$ being a time-dependent \textit{parameter} of $\bm Q(t)$. The resulting \textit{tangent nonhomogeneous Markov process} has the following transient \textit{stochastic matrix}
\begin{equation}\label{StochasticM}
\bm S(t,t+\Delta t)\defi\bm I+\bm Q(t)\Delta t\,.\
\end{equation}
In line with the macroscopic description (Eq.\eqref{Normalize}), the initial condition $\bm P(t_0)$ as well as the rate matrix $\bm Q(\bm P(t),t)$ are by definition exactly the same for all $N$ individuals.
This assumption corresponds  \textit{ad verbum} to fix a constant average number of contacts and other interaction parameters between persons per unit time. Given this, there is an implicit assumption of statistical independence among the $N$ individuals. In an adiabatic system, this is equivalent to letting $N$ particles following $N$ independent Brownian motions. Therefore, this macro-description is emerging from the micro-behavior of individuals interacting according $N$ independent Brownian motions, and the virus is spreading according a simple diffusive process\footnote{At the micro-level, the spreading of the virus can be better described through a stochastic branching process. The simple diffusion process can be regarded as the result of a coarse-graining.}. Consequently, a macro-random scenario of an epidemic can be obtained via simulating $N$ independent and identically distributed processes from Eq.\eqref{PMasterEqDiff}.

In contrast with this macro-description, one could adopt a topological structure of the interactions between different individuals. This is generally done by including an adjacency operator which accounts for the different structure of the interactions among individuals (e.g., including mobility information, or considering the presence of superspreaders). As a consequence, each individual (or group of individuals) has a different average number of contacts and different interaction parameters. This leads to an heterogeneous compartmental model, which is inevitable dependent on a specific geographical area or social system. In this study, we focus on the general transmission trend of large regions, so that the trends can be more easily extrapolated and interpreted. Therefore, the simple macro-description is adopted. It is a specific choice which leads to a novel entropy-based measure to macroscopically compare the epidemic scenarios in different regions.

\subsection{Time-dependent parameter model}
\noindent We assume the ``correct'' model of $\bm Q(\bm P(t),t)$ cannot be discovered, and $\bm Q(\bm P(t),t)$ is replaced by a parametric model with a set of parameters $\bm\alpha(t)$, i.e.,
\begin{equation}\label{ParametricQ}
\bm Q(\bm P(t),t)\approx{\bm Q}(\bm P(t),t;\bm\alpha(t))\,.\
\end{equation}
Let $\alpha(t)$ represent an arbitrary component of $\bm\alpha(t)$. A generic approach to parameterize $\alpha(t)$ is to consider an expansion of the following form
\begin{equation}\label{ExpandAlpha}
\alpha(t)=\sum_{i=0}^{I}w_i\psi_i(t)\,,\
\end{equation}
where $w_i$ are coordinates of basis functions $\psi_i(t)$. A popular choice for the basis function is the orthogonal polynomials, e.g., Legendre/Hermite/Laguerre/Chebyshev polynomials. An issue with orthogonal polynomial basis is that it may require high-order terms to represent a complex function, and consequently this leads to over-fitting and implausible extrapolations. A powerful alternative is to use adaptive basis functions with the form
\begin{equation}\label{ExpandAlphaAdapt}
\alpha(t)=\sum_{i=0}^{I}w_{i}\psi_i(t,\bm w'_{i})\,,\
\end{equation}
where $\bm w'_i$ are parameters of the adaptive basis. The benefit of using Eq.\eqref{ExpandAlphaAdapt} instead of Eq.\eqref{ExpandAlpha} is that a parsimonious representation can be formulated, at the cost of introducing additional parameters in bases. An attractive choice for the adaptive basis $\psi_i(t,\bm w_i)$ is the sigmoid function, i.e.,
\begin{equation}\label{AlphaSigmoid}
\psi_i(t,\bm w'_i)=\frac{1}{1+\exp(w'_{i1}-w'_{i2}t)}\,.\
\end{equation}
The theoretical justification of using Eq.\eqref{AlphaSigmoid} in Eq.\eqref{ExpandAlphaAdapt} is the \textit{universal approximation theorem} \cite{ApproxTheorem}, and the resulting parametric function is in fact a feed-forward neural network with a single hidden layer.

In addition, the initial condition of Eq.\eqref{PMasterEqDiff} is unknown, and we parameterize $\bm P(t_0)$ by $\bm\beta=[\beta_1,...,\beta_{M-1}]$ (recall that $M$ is the number of compartments). A natural parametrization of $\bm P(t_0)$ is
\begin{equation}\label{ParametricP0}
\bm P(t_0)=\left[\beta_1,...,\beta_{M-1},1-\sum_{m=1}^{M-1}\beta_m\right]\tr\,,\
\end{equation}
where $\beta_m$ are nonnegative and subjected to the linear constraint $\sum_{m=1}^{M-1}\beta_m\in[0,1]$. Note that $\bm\beta$ is time-independent in the sense that the starting time point can be fixed. Therefore, the full parameter set of the epidemic model is written as $\bm\theta:=\left\lbrace\bm w, \bm w', \bm\beta\right\rbrace$.

\subsection{Model calibration}\label{Sec:ModCal}

\noindent The goal of model calibration is to find the optimal $\bm\theta$ using real observation. We let $\bm{\mathcal{D}}$ denote the dataset of observations collected for an epidemic up to some reference time point. The dataset $\bm{\mathcal{D}}$ is composed by discrete measures on the number of persons in each observable compartment (e.g., infected, recovered, and dead), and $\bm{\mathcal{D}}$ is a matrix of dimension $M_o\times T$, where $M_o$ denotes the number of observable compartments, and $T$ denotes the number of observed unit time (e.g., days).

The likelihood function $\mathcal{L}(\bm{\mathcal{D}}|\bm\theta)$ measures the probability of observing $\bm{\mathcal{D}}$ given the model specified by $\bm\theta$. Using Bayes rule on $\bm\theta$, we have
\begin{equation}\label{Bayes}
\pi(\bm\theta|\bm{{\mathcal{D}}})\propto \mathcal{L}(\bm{{\mathcal{D}}}|\bm\theta)\pi(\bm\theta)\,,
\end{equation}
where $\pi(\bm\theta|\bm{{\mathcal{D}}})$ is the \textit{posterior} distribution of $\bm\theta$ conditional on the observed dataset $\bm{{\mathcal{D}}}$, and $\pi(\bm\theta)$ is the \textit{prior} distribution of $\bm\theta$. The major challenge of using Eq.\eqref{Bayes} in practice is to sample from the posterior, and typically this can be handled by advanced Markov Chain Monte Carlo methods.

The likelihood $\mathcal{L}(\bm{\mathcal{D}}|\bm\theta)$  may depend both on the observation error and the inherent variability of the epidemic model. Even by setting the observation error to zero, for any specified $\bm\theta$ the prediction from the model is still random. If the accumulated numbers are of interest, e.g., the total number of recovered, for a large population size the variability in the prediction is expected to be small. Specifically, in a multinomial model the marginal coefficient of variation is proportional to $1/\sqrt{NP_m(t)}$. However, at the same time, the model prediction can be extremely sensitive to $\bm\theta$, and an almost negligible perturbation due to the (albeit small) randomness of $\bm\theta$ may lead to noticeably different predictions. Therefore, the Bayesian analysis is meaningful with or without the observation error.

To formulate the likelihood function, we first denote an individual (directed) random walk among various states as a boolean operator $\bm Y^{(n)}=[\bm y^{(n)}_1,...,\bm y^{(n)}_j,...,\bm y^{(n)}_T]$, where $n\in[1,...,N]$ and $j\in[ 1,...,T]$ such that $t_{j+1}-t_j = 1$ [unit time]\footnote{Although Eq.\eqref{PMasterEqDiff} is continuous in time, the observations are recorded in discrete time points, therefore here $t$ is discretized.}. The vectors $\bm y^{(n)}_j$ (of dimension $M_o\times 1$) represent the state of the $n$ person at time $t_j$. Therefore the components are all zero with exception of the current state, which takes the value of one. The joint probability density function of $\bm Y^{(n)}$, denoted by $f(\bm Y^{(n)}|\bm\theta)$ is readily available from the governing equation Eq.\eqref{PMasterEqDiff}. Next, we note that the observation $\bm{\mathcal{D}}$ represents a collective scenario of the $N$ independent (under the assumptions of the macro-model) Markov processes $\bm y^{(n)}_j$.

Therefore, by brute force, the (observation-error-free) likelihood function has the following form
\begin{equation}\label{Likelihood}
\mathcal{L}(\bm{\mathcal{D}}|\bm\theta)=\sum_{y^{(n)}_{m,t}}\left[\mathbbm{1}\left(\sum_{n}\bm Y^{(n)} = \bm{\mathcal{D}\right)}\prod_{n} f\left(\bm Y^{(n)}|\bm \theta\right)  \right]
\end{equation}
where $\mathbbm{1}(\cdot)$ is an indicator function, and $\sum_{y^{(n)}_{m,t}}$ is a $M_o^{T\times N}$ fold summation.  This brute force summation contains impossible paths that, however, are naturally excluded by the indicator function. Observe that this likelihood is fundamentally different from the deterministic compartmental models based on proportions rather than individual probabilities. Moreover, it is also different from the classical binomial (and related) likelihood approaches (used in direct Gillespie's methods). Eq.\eqref{Likelihood} is clearly computationally intractable.

To formulate a computationally tractable likelihood function, we use the Markovian property and rewrite $\mathcal{L}(\bm{\mathcal{D}}|\bm\theta)$ as
\begin{equation}\label{LikelihoodMarkov}
\mathcal{L}(\bm{\mathcal{D}}|\bm\theta)=\mathcal{L}(\bm{\mathcal{D}}(t_0)|\bm\theta)\prod_{j=1}^{T}\mathcal{L}(\bm{\mathcal{D}}(t_j)|\bm{\mathcal{D}}(t_{j-1});\bm\theta)\,,
\end{equation}
where $\bm{\mathcal{D}}(t_j)$ denotes the observation at time point $t_j$. The first term $\mathcal{L}(\bm{\mathcal{D}}(t_0)|\bm\theta)$ can be easily computed from a multinomial distribution with probability vector $\bm P(t_0)$. The specific expression of $\mathcal{L}(\bm{\mathcal{D}}(t_j)|\bm{\mathcal{D}}(t_{j-1});\bm\theta)$ varies with the adopted epidemic model, yet it is typically in the multinomial form. All the ingredients to compute $\mathcal{L}(\bm{\mathcal{D}}(t_j)|\bm{\mathcal{D}}(t_{j-1});\bm\theta)$ are included in the marginal distribution $\bm P(t)$, and the stochastic matrix $\bm S(t_j,t_{j+1}|\bm\theta)$ expressed as
\begin{equation}\label{StochasticMatrixtj}
\bm S(t_j,t_{j+1}|\bm\theta)=\exp\left(\int_{t_j}^{t_{j+1}}\bm Q(\tau|\bm\theta)\,d\tau\right)\,.\
\end{equation}
Observe that due to the discretization of a continuous time Markov process into a discrete time Markov process, the matrix $\bm S(t_j,t_{j+1}|\bm\theta)$ is ``less sparse" than $\bm Q(t|\bm\theta)$. For example, in a finite time interval, the impossible event $2\to4$ in matrix $\bm Q$ may have a finite probability of occurring in matrix $\bm S$ (through, e.g. $2\to3\to4$). In fact, Eq.\eqref{StochasticMatrixtj} can be interpreted as the result of applying Eq.\eqref{StochasticM} infinite times within the integration interval.

For a simple illustration of concept in constructing the likelihood function, we consider a two state system where state 1 can either move to state 2 or stay still, while state 2 can only stay still. We assume $\bm{\mathcal{D}}(t_{j-1})$ records $[100,50]$ in occupations of states 1 and 2 (for a total of 150 Markov chains), and $\bm{\mathcal{D}}(t_{j})$ records $[90,60]$. Given the aforementioned transition structure, we know 10 out of 100 chains at $t_{j}$ moves from state 1 to state 2. Therefore, the likelihood $\mathcal{L}(\bm{\mathcal{D}}(t_j)|\bm{\mathcal{D}}(t_{j-1});\bm\theta)$ is simply the binomial ${{100}\choose{10}}P_{1\to2}^{10}P_{1\to1}^{90}$, where the transition probability $P_{i\to j}$ can be directly read from Eq.\eqref{StochasticMatrixtj}. One may not be able to observe the populations in all compartments, in this case the total probability theorem can be used to integrate the unobservable states out (see Section \ref{Application} for an example).

\subsection{Addressing data unreliability}
\noindent The likelihood function introduced above only considers the inherent stochastic variability of the model. In reality, on top of the inherent stochastic variability, the underlying errors/uncertainties of a reported dataset involve multiple alternative sources. A rigorous way to treat such unreliability of reported data is to introduce a distribution assumption on the error $\bm\epsilon$, and the likelihood function can be written as
\begin{equation}\label{ErrorLikelihood}
\mathcal{L}(\bm{{\mathcal{D}}}|\bm\theta)=\int_{\bm\epsilon\in\Omega_{\bm\epsilon}}\mathcal{L}(\bm{{\mathcal{D}}}|\bm\theta,\bm\epsilon)\pi(\bm\epsilon)\,d\bm\epsilon\,,
\end{equation}
where $\mathcal{L}(\bm{{\mathcal{D}}}|\bm\theta,\bm\epsilon)$ is the likelihood with a specified error, $\pi(\bm\epsilon)$ is the probability distribution of the error, and $\Omega_{\bm\epsilon}$ represents the feasible domain of the error. Note that in general $\bm\epsilon$ represents a set of discretized stochastic processes. Apart from the technical challenge of integrating the high-dimensional Eq.\eqref{ErrorLikelihood}, the major challenge of incorporating the error is the specification of $\pi(\bm\epsilon)$. Clearly, an assumption on $\pi(\bm\epsilon)$ would reshape the likelihood function towards the shape of $\pi(\bm\epsilon)$, and an inappropriate assumption would generate artificial and even misleading transmission properties. Therefore, we adopt an indirect path to incorporate the unreliability of reported data. Specifically, we apply a kernel function $\kappa(\cdot)$ to the original error-free likelihood function, i.e.,
\begin{equation}\label{KernelLikelihood}
\hat{\mathcal{L}}(\bm{{\mathcal{D}}}|\bm\theta)=\kappa\left({\mathcal{L}}(\bm{{\mathcal{D}}}|\bm\theta)\right)\,.
\end{equation}
The kernel function is selected to ``flatten'' the likelihood function so that the unreliability in the reported data can be, to some extent, captured. In this study, we consider an exponential kernel, and Eq.\eqref{KernelLikelihood} is rewritten as
\begin{equation}\label{GaussianKernelLikelihood}
\hat{\mathcal{L}}(\bm{{\mathcal{D}}}|\bm\theta)=\exp\frac{\log{\mathcal{L}}(\bm{{\mathcal{D}}}|\bm\theta)}{n_{\epsilon}}\,,
\end{equation}
where $n_{\epsilon}>1$ is a scaling factor. Clearly, if $n_{\epsilon}=1$, $\hat{\mathcal{L}}(\bm{{\mathcal{D}}}|\bm\theta)$ is identical to the original likelihood ${\mathcal{L}}(\bm{{\mathcal{D}}}|\bm\theta)$; and if $n_{\epsilon}\to\infty$, $\hat{\mathcal{L}}(\bm{{\mathcal{D}}}|\bm\theta)$ approaches uniform.

Instead of a specific error distribution, in practice it is more likely to have a crude idea on the possible magnitude of the errors in the reported dataset. For a further simplification, we focus on the errors in the infected cases, since the causal structure of infected and recovered/dead would let the errors in infected eventually flow into recovered/dead. Therefore, the question left is to relate ``the magnitude of errors in infected cases'' to the $n_{\epsilon}$ in Eq.\eqref{GaussianKernelLikelihood}. It turns out, as a consequence of a sequence of qualitative reasoning, a reasonable choice of $n_{\epsilon}$ is to let
\begin{equation}\label{neps}
n_{\epsilon}\propto\frac{\Delta_\epsilon^2}{\Delta_{infected}}\,,
\end{equation}
where $\Delta_{infected}$ represents the maximum increment of infected, and $\Delta_\epsilon$ represents the possible error in the maximum increment of infected. Note that Eq.\eqref{neps} is proposed as a crude guidance for setting the magnitude of $n_{\epsilon}$. The reasoning of Eq.\eqref{neps} is described as follows. i) In Eq.\eqref{GaussianKernelLikelihood}, if ${\mathcal{L}}(\bm{{\mathcal{D}}}|\bm\theta)$ is Gaussian, the effect of applying $1/n_{\epsilon}$ is to introduce a scaling factor of $n_{\epsilon}$ to the covariance of ${\mathcal{L}}(\bm{{\mathcal{D}}}|\bm\theta)$. ii) The likelihood ${\mathcal{L}}(\bm{{\mathcal{D}}}|\bm\theta)$ is a product of multinomial kernels (see Eq.\eqref{LikelihoodMarkov}), which can be approximated by Gaussian with the maximum variance (of the infected compartment) in the size of $\Delta_{infected}$. iii) Eq.\eqref{neps} is obtained as one assumes the scaled variance (scaled by factor $n_{\epsilon}$) has a similar magnitude as $\Delta_{\epsilon}^2$ \footnote{This assumption implies if $\Delta_\epsilon=\sqrt{\Delta_{infected}}$, $n_{\epsilon}$ is $1$. In other words, if the error is in the size of the standard deviation of multinomial distribution, one cannot tell it is error or inherent stochastic variability.}. For example, if one has a crude idea that the error of infected can be 30\% of the reported infected, using Eq.\eqref{neps} one could set $n_{\epsilon}\propto0.09\Delta_{infected}$.

\section{Entropy as a global transmission metric}\label{Sec:Entropy}

\noindent The key feature of the proposed stochastic model is that entropy-based transmission measures can be naturally developed. Specifically, for a discretized time grid $\lbrace t_j,j=1,...,T\rbrace$ and stochastic matrix $\bm S(t_j,t_{j+1})$ (Eq.\eqref{StochasticMatrixtj}), we consider the Shannon entropy rate expressed as
\begin{equation}\label{Entropy}
\mathcal{H}(t_j|t_{j-1})=-\sum_{m=1}^M\sum_{n=1}^MP_n(t_{j-1})S_{m,n}(t_{j-1},t_{j})\log(S_{m,n}(t_{j-1},t_{j}))\,.\
\end{equation}
For $j=0$, $\mathcal{H}(t_0|t_{-1})\equiv\mathcal{H}(t_0)=-\sum_{m=1}^{M}P_m(t_0)\log P_m(t_0)$. Recall that the marginal distribution $\bm P(t)$ and the stochastic matrix $\bm S(t_j,t_{j+1})$ vary with the initial condition $\bm P(t_0)$. Therefore, Eq.\eqref{Entropy} and $\mathcal{H}(t_0)$ should be averaged over the posterior distribution of the initial condition (obtained from the Bayesian analysis). In evaluation of Eq.\eqref{Entropy}, the convention $0\log0\equiv0$ is adopted.

In a homogeneous Markov process, the entropy rate is constant, and one has the important theoretical result $\lim_{T\to\infty}\frac{1}{T}\mathcal{H}(t_0,t_1,...,t_T)=\mathcal{H}(t_1|t_0)$. In the proposed epidemic model, the Markov process is nonlinear and nonhomogeneous. Therefore, the evolution of the entropy rate $\mathcal{H}(t_j|t_{j-1})$ within a specified duration should be considered, and they characterize the evolution of the degree of disorder.

Using the Markovian property of the epidemic model in conjunction with the additive property of entropy\footnote{Note that the entropy is additive when the underlying distributions are statistically independent, and in Markov processes the consecutive conditional distributions are independent.}, the entropy $\mathcal{H}(t_0,t_1,...,t_T)$ has the concise form
\begin{equation}\label{EntropyTotal}
\mathcal{H}(t_0,t_1,...,t_T)=\sum_{j=0}^T\mathcal{H}(t_j|t_{j-1})\,.\
\end{equation}
The entropy $\mathcal{H}(t_0,t_1,...,t_T)$ is a scalar and it provides a global measure on the total degree of disorder for an epidemic scenario. An important feature (shared by the reproductive ratio) of the entropy rate and the total entropy is that they are quantitatively comparable across different regions. This is because the entropy-based measures are associated with the statistically averaged individual, which is similar to measuring the mean-field approximation of the complex epidemic dynamics system.

Qualitative speaking, a large $\mathcal{H}(t_0,t_1,...,t_T)$ may be contributed by: i) a large pulse-like $\mathcal{H}(t_j|t_{j-1})$, i.e. the entropy rate reaches high values but stays (in high values) for a short period; ii) a moderate flat $\mathcal{H}(t_j|t_{j-1})$, i.e. the entropy rate evolves with moderate values for a long period. In an epidemic scenario, a large pulse-like evolution of the entropy rate implies that the virus reaches a significant proportion of population but damped out (through the accumulation of recovered/dead) fast, and a flat evolution implies that the epidemic spreads in a moderate severe state for a long time. To quantitatively analyze if the entropy rate evolution is pulse-like or flat, we introduce a concentration measure to $\mathcal{H}(t_j|t_{j-1})$. Specifically, we again adopt the concept of Shannon entropy such that the concentration measure of $\mathcal{H}(t_j|t_{j-1})$ is defined as the inverse of the Shannon entropy of the normalized $\mathcal{H}(t_j|t_{j-1})$, i.e.
\begin{equation}\label{Concentration}
\mathcal{C}(\mathcal{H})=\frac{1}{\mathcal{H}(\bar{\mathcal{H}}(t_{j}|t_{j-1}))}=-\frac{\mathcal{H}(t_0,t_1,...,t_T)}{\sum_{j=0}^T\mathcal{H}(t_{j}|t_{j-1})(\log{\mathcal{H}(t_{j}|t_{j-1})}-\log{\mathcal{H}(t_0,t_1,...,t_T))}}\,,\
\end{equation}
where $\bar{\mathcal{H}}(t_{j}|t_{j-1}))=\mathcal{H}(t_{j}|t_{j-1}))/\mathcal{H}(t_0,t_1,...,t_T)$ is the normalized $\mathcal{H}(t_j|t_{j-1})$. Note that the total entropy $\mathcal{H}(t_0,t_1,...,t_T)$ appears in Eq.\eqref{Concentration} as the normalizing constant of $\mathcal{H}(t_j|t_{j-1})$ (when $\mathcal{H}(t_j|t_{j-1})$ is normalized into a probability mass function). Also note that the Shannon entropy instead of the variance-based measures is adopted since a large variance does not necessarily reflect a large dispersion (e.g., a mixture model with highly concentrated component densities could produce a large variance).

In this paper, we propose the entropy rate $\mathcal{H}(t_j|t_{j-1})$, the entropy $\mathcal{H}(t_0,t_1,...,t_T)$ and the concentration factor $\mathcal{C}(\mathcal{H})$ as complements to the conventional reproductive ratio. \ref{Append:Entropy} illustrates various attractive features of the entropy-based measures.
In practice, instead of computing entropy-based measures for the original distribution vector $\bm P$ and the stochastic matrix $\bm S$, one may need to reshape $\bm P$ and $\bm S$ to obtain measures of different scales. For example, a typical epidemic model may involve the states of recovered and dead. Naturally, one would prefer the scenario of ``a large recovery probability and a small death probability" over ``a large death probability and a small recovery probability." However, Eq.\eqref{Entropy} or Eq.\eqref{EntropyTotal} does not differentiate between recovery and death, and the aforementioned two scenarios can have exactly the same entropy (rate). The conventional reproductive ratio measure has the same issue. To let the entropy-based measures incorporate the concept of ``high recovery probability is preferable over high death probability'', one could reshape the distribution vector $\bm P$ and the stochastic matrix $\bm S$ by merging the recovery state with the infected state. Consequently, the entropy (rate) of the reshaped system would diminish the contribution from the recovered state and highlight the contribution from the dead state (see \ref{Append:Entropy} for an example).

\section{Application to COVID-19}\label{Application}
\noindent In light of the general framework introduces in Section \ref{Framework}-\ref{Sec:Entropy}, this Section introduces a simple modification of the SEIR model with the exposed being also contagious.

\subsection{Modified SEIR}
\noindent The modified SEIR has a 5-dimensional probability state vector $\bm P(t)$ described as follows.
\begin{itemize}
  \item $P_1(t)$: the (instantaneous probability of being) susceptible.
  \item $P_2(t)$: the exposed.
  \item $P_3(t)$: the infected.
  \item $P_4(t)$: the recovered.
   \item $P_5(t)$: the dead.
\end{itemize}
The rate matrix $\bm Q(\bm P(t),t)$ is written as
\begin{equation}\label{MatQSEIR}
\bm Q(\bm P(t),t)=\begin{bmatrix}
-(\alpha_1(t)P_2(t)+\alpha_2(t)P_3(t)) & 0 & 0 & 0 & 0 \\
\alpha_1(t)P_2(t)+\alpha_2(t)P_3(t) & -\alpha_3(t) & 0 & 0 & 0 \\
0 & \alpha_3(t) & -(\alpha_4(t)+\alpha_5(t)) & 0 & 0 \\
0 & 0 & \alpha_4(t) & 0 & 0 \\
0 & 0 & \alpha_5(t) & 0 & 0 \\
\end{bmatrix}\,,\
\end{equation}
where $\bm\alpha(t)=[\alpha_1(t),...,\alpha_5(t)]$ are non-negative parameters to be calibrated. Note that here for generality we write every parameter as time-dependent; however, in practice it is typically sufficient to set only a few of them as time-dependent.

\subsection{Likelihood function}
\noindent The likelihood function can be derived as a simple application of the concepts introduced in Section \ref{Sec:ModCal}. First, we introduce a compound state, denoted by $1\lor2$, to represent the state of being in either susceptible or exposed. The most important property of the compound state $1\lor2$ is that it is an observable, i.e., if an individual is not at the state of infected nor at the state of recovered/dead, it is in the compound state. We let $P_{1\lor2\to m}(t_j,t_{j+1})$ represent the transition probability from the compound state at $t_j$ to other states $m$, $m=3$ (infected), $4$ (recovered), $5$ (dead) at $t_{j+1}$. Using the total probability theorem, $p_{1\lor2\to m}(t_j,t_{j+1})$ can be expressed as
\begin{equation}\label{CompoundTransit}
P_{1\lor2\to m}(t_j,t_{j+1})=\frac{P_1(t_j)}{P_1(t_j)+P_2(t_j)}P_{1\to m}(t_j,t_{j+1})+\frac{P_2(t_j)}{P_1(t_j)+P_2(t_j)}P_{2\to m}(t_j,t_{j+1})\,,\
\end{equation}
where $P_1(t_j)$ and $P_2(t_j)$ are solution of Eq.\eqref{PMasterEqDiff}, and $P_{1\to m}$ and $P_{2\to m}$ can be obtained from Eq.\eqref{StochasticMatrixtj}.
Next, we arrange the dataset vector $\bm{\mathcal{D}}(t_j)$ in the form $\bm{\mathcal{D}}(t_j)=\left[\mathcal{D}_{1\lor2}(t_j),\mathcal{D}_{3}(t_j),\mathcal{D}_{4}(t_j),\mathcal{D}_{5}(t_j)\right]$ to respectively represent the instantaneous number of compound state, instantaneous number of infected, accumulative number of recovered, and accumulative number of dead. Let $\Delta\bm{\mathcal{D}}(t_j,t_{j+1})\defi|\bm{\mathcal{D}}(t_{j+1})-\bm{\mathcal{D}}(t_j)|$ represent the absolute difference between two consecutive dataset vectors.

Before introducing the likelihood function, we introduce an additional assumption that the $\Delta\mathcal{D}_{1\lor2}(t_j,t_{j+1})$ number of Markov chains all transit to the state 3 (the infected). This assumption can be always (made) correct since: i) if $t_j$ is sufficiently close to $t_{j+1}$, naturally one cannot jump to the recovered/dead state from susceptible/exposed; ii) if $t_{j+1}-t_j$ is large, one could re-mesh the time scale and perform interpolation on the dataset, so that $t_j$ can always be close to $t_{j+1}$ \textit{by construction}.
Finally, the conditional likelihood function $\mathcal{L}(\bm{\mathcal{D}}(t_{j+1})|\bm{\mathcal{D}}(t_j);\bm\theta)$ can be written as
\begin{equation}\label{LiklihoodSEIR}
\mathcal{L}(\bm{\mathcal{D}}(t_{j+1})|\bm{\mathcal{D}}(t_j);\bm\theta)=\mathcal{L}_1\mathcal{L}_2\,,\
\end{equation}
where
\begin{equation}\label{LiklihoodSEIR1}
\mathcal{L}_1=\frac{\mathcal{D}_{1\lor2}!}{(\mathcal{D}_{1\lor2}-\Delta\mathcal{D}_{1\lor2})!\Delta\mathcal{D}_{1\lor2}!0!0!}(P_{1\lor2\to 1\lor2})^{\mathcal{D}_{1\lor2}-\Delta\mathcal{D}_{1\lor2}}(P_{1\lor2\to 3})^{\Delta\mathcal{D}_{1\lor2}}(P_{1\lor2\to4})^{0}(P_{1\lor2\to5})^{0}
\end{equation}
and
\begin{equation}\label{LiklihoodSEIR2}
\mathcal{L}_2=\frac{\mathcal{D}_{3}!}{(\mathcal{D}_{3}-\Delta\mathcal{D}_{4}-\Delta\mathcal{D}_{5})!\Delta\mathcal{D}_{4}!\Delta\mathcal{D}_{5}!}(P_{3\to 3})^{\mathcal{D}_{3}-\Delta\mathcal{D}_{4}-\Delta\mathcal{D}_{5}}(P_{3\to4})^{\Delta\mathcal{D}_{4}}(P_{3\to5})^{\Delta\mathcal{D}_{5}}
\end{equation}
In Eq.\eqref{LiklihoodSEIR1} and Eq.\eqref{LiklihoodSEIR2}, the notations are simplified to drop $t_j$, $t_{j+1}$ and $\bm\theta$. To avoid possible ambiguity, the simplification rules are: $\mathcal{D}_{m}\equiv\mathcal{D}_{m}(t_j)$, $\Delta\mathcal{D}_{m}\equiv\Delta\mathcal{D}_{m}(t_j,t_{j+1})$, $P_{m}\equiv P_{m}(t_j|\bm\theta)$, and $P_{m\to m'}\equiv P_{m\to m'}(t_j,t_{j+1}|\bm\theta)$. Substituting Eq.\eqref{LiklihoodSEIR}, Eq.\eqref{LiklihoodSEIR1} and Eq.\eqref{LiklihoodSEIR2} into Eq.\eqref{LikelihoodMarkov}, one obtains the complete likelihood function.

\subsection{Transmission measures}\label{Sec:Transmission}
\noindent To obtain the entropy-based measures, we reshape the 5-dimensional vector $\bm P$ into $[P_{1\lor2},P_{3\lor4},P_{5}]\tr$, and the corresponding stochastic matrix $\bm S$ is also reshaped (via the total probability theorem) accordingly. The reason to consider the compound state $1\lor2$ is because as a whole the state $1\lor2$ is an observable, therefore the possible errors in identifying the exposed can be marginalized out. The reason to consider the compound state $3\lor4$ is discussed in Section \ref{Sec:Entropy}, and as a result the concept ``high recovery probability is preferable over high death probability" is correctly incorporated.

In addition to the entropy-based measures, using the next-generation matrix approach \cite{MathematicalEpidemiology}, the instantaneous reproductive ratio of the modified SEIR model can be defined as
\begin{equation}\label{R0SEIR}
R_0(t)\defi\frac{\alpha_1(t)}{\alpha_3(t)}+\frac{\alpha_2(t)}{\alpha_4(t)+\alpha_5(t)}\,.\
\end{equation}
Note that the instantaneous reproductive ratio $R_0(t)$ can be understood as the basic reproductive ratio of a tangent model defined as a constant model with parameters $\bm\alpha$ equal to the instantaneous parameters $\bm\alpha(t)$ at the reference time point $t$ .

\subsection{Modelling and computational details}
\noindent As aforementioned, the whole parameter set $\bm\theta$ not only involves parameters of the rate matrix $\bm Q(\bm P(t),t)$, i.e., $\bm\alpha(t)$ (represented by $\bm w, \bm w'$ of basis functions), but also parameters to represent the initial state, i.e., $\bm\beta$. In the modelling practice, except $\alpha_3$, which is related to the mean incubation period, we calibrate all the other parameters (including the initial conditions) with Bayesian analysis. The mean incubation period, which is $1/\alpha_3$ in the model, is reported in various previous studies \cite{Incubtation}\cite{WHO}, and typically it is around 5 and in the range of $[3,7]$ days. Therefore, we set $1/\alpha_3$ as an epistemic random variable within $[3,7]$. The time-dependent parameters are modeled with sigmoid basis functions. The number of function basis for each parameter is determined in an additive manner. Specifically, we start with constant $\alpha$ and iteratively increase the number of basis functions until the variation in likelihood function value (Eq.\eqref{LikelihoodMarkov}) or BIC index becomes small.

The Gibbs sampling with a uniform proposal distribution for each component of $\bm\theta$ is adopted to sample from the posterior distribution. The step size of the Gibbs sampling is adaptively tuned using the acceptance rate of the Markov chain \cite{haario2001adaptive}\cite{WANG201951}. The seed samples for the Gibbs sampler are selected in the neighborhood of the posterior mode. This is obtained by sequential Monte Carlo method \cite{moral2006sequential}\cite{C2012Sequential} combined with deterministic trust region optimization \cite{LevenbergAlgorithm}\cite{advancestrustregion}.

\section{Modelling results on real datasets of COVID-19}\label{NumericalTest}
\subsection{Data sets}
\noindent For the studied regions, the time series of the populations of infected, recovered and dead during January to May, 2020 are used in model calibration. The data is collected from WHO, European CDC and Chinese CDC \cite{WHO}\cite{CDCEurope}\cite{CDCChina}. The regions considered in this study include: Hubei province, South Korea, Italy, Germany, Spain and France\footnote{We planned to include also UK; however, the data on the recovered patients are not available yet (unfortunately).}. For each region, the population size $N$ is fixed to the most recent value reported by Worldometer \cite{Populatoin}. We choose these countries/regions because they have the same order of population size (this is irrelevant to the entropy-based measure which is $N$-independent), they applied different containment strategies, and they represent different cultures. Moreover, at the time of writing of this article the peak of the epidemic waves is passed. A complete and thorough analysis of a large number of regions is out of the scope of the current study. In fact, here we focus primarily on the model and metric definition and their use. 
\subsection{Data correction}
\begin{figure}[H]
    \centering
    \includegraphics[scale=0.75]{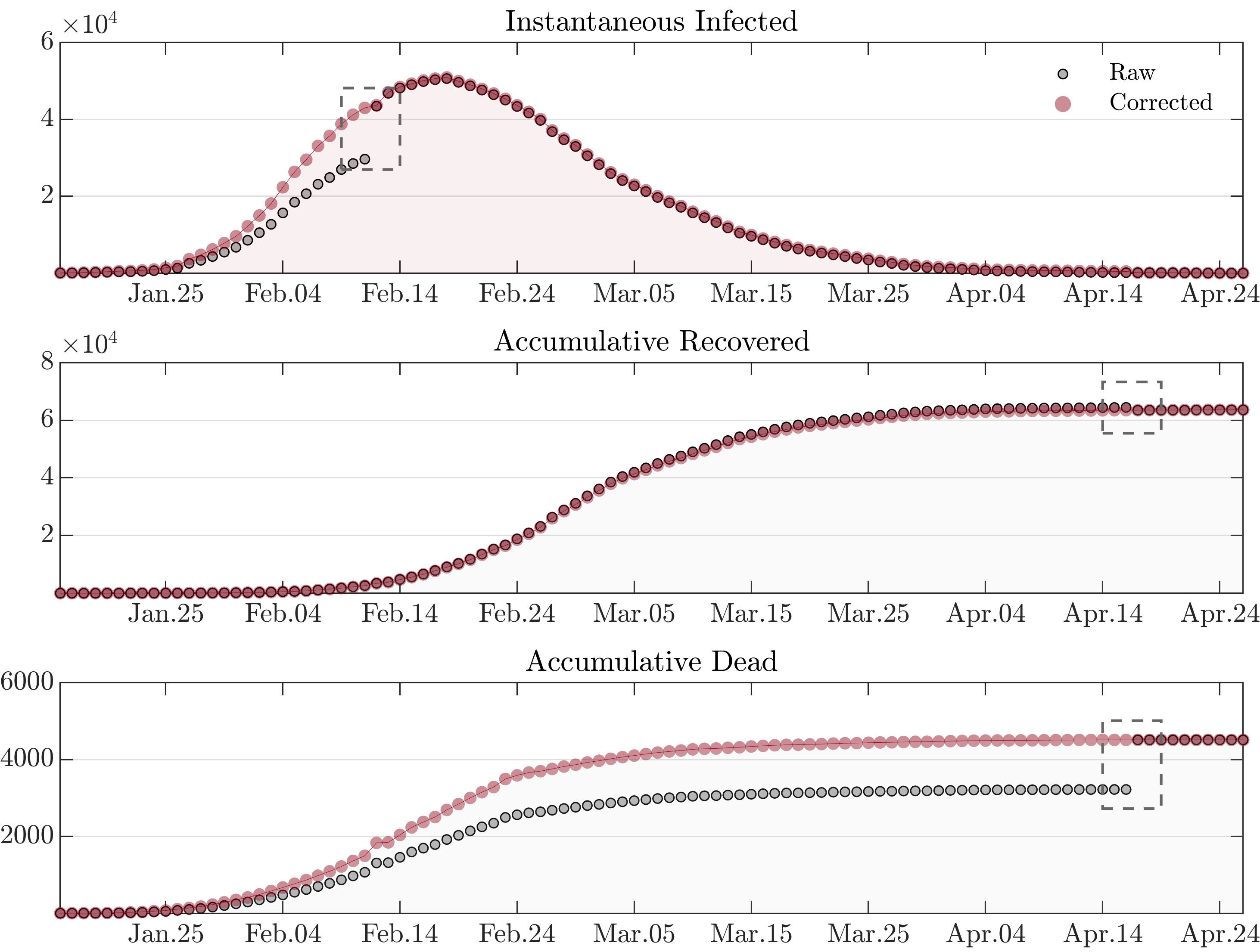}
    \caption{\textbf{Raw and corrected datasets of Hubei province}. \textit{There are two policy changes regarding the dataset: i) in February 12, 2020, the diagnosis criterion was temporarily relaxed, and as a result there is an artificial jump in the number of infected; ii) in April 17, 2020, the cumulative number of infected and dead are altered by a constant jump, and the cumulative number of recovered is altered by a constant drop. The jumps/drops are marked by a rectangular in the figure. The populations of infected, recovered and dead are corrected using Eq.\eqref{Postprocess}. Note that for infected the correction is made on cumulative numbers, and then the instantaneous infected is obtained by subtracting the accumulative recovered and dead}.}
    \label{Fig:Data}
\end{figure}

\noindent Due to abrupt counting policy changes and various corrections, the COVID-19 datasets for Hubei, Spain and France not only violate the smoothness assumption\footnote{Due to the inherent stochastic variability, the random samples drawn from the model is naturally non-smooth. However, here the ``violence of smoothness assumption'' indicates that an \textbf{artificial manipulation} of data will corrupt the smoothness of propagator $\bm Q$.} of the proposed modeling framework, but also contradict the fundamental fact that the accumulative number can only be non-decreasing. Therefore, the datasets must be corrected. It is obvious that the cluster/jump of data has a missing information, which is the (correct) time of occurrence. To obtain a consistent dataset we fill this missing information by using the expected time of occurrence with respect to the distribution of the previous events. Since the dataset is recorded daily, marginally they form a multinomial distribution along the discrete time axis. It follows that the missing time information can be filled by using the daily expected number of events. Specifically, let $t_J$ represent the time point when the jump/drop happens (for a specified compartment), and let $\Delta\mathcal{D}_J$ represent the magnitude of the data jump/drop. We perform a postprocess of the dataset expressed as follows
\begin{equation}\label{Postprocess}
\mathcal{D}(t_i)\leftarrow\left(1+\frac{\Delta\mathcal{D}_J}{\mathcal{D}(t_{J})}\right)\mathcal{D}(t_i)\,,\
\end{equation}
where $t_i=t_0,t_1,...,t_{J}$, and $\mathcal{D}(t_i)$ represents the cumulative\footnote{In Eq.\eqref{Postprocess} $\mathcal{D}(t_i)$ has to be the cumulative number instead of instantaneous one.} number at $t_i$. Note that $\Delta\mathcal{D}_J$ could be negative. For an illustration of the correction, Figure \ref{Fig:Data} shows the raw and the corrected datasets for Hubei province.

\subsection{The overall epidemic dynamics of various regions}
\begin{figure}[H]
    \centering
    \includegraphics[scale=0.7]{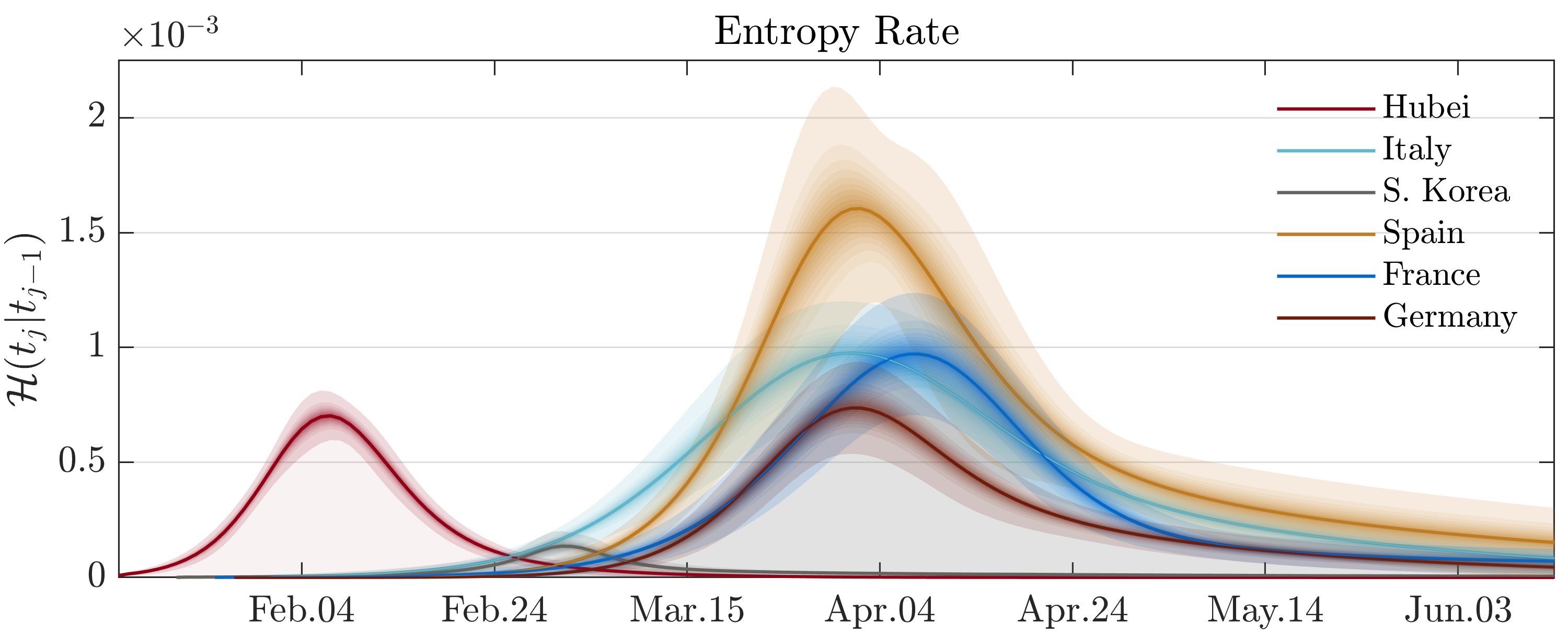}
    \caption{\textbf{The entropy rates for various regions}. \textit{The figure shows the temporal evolution of entropy rate for various regions. The solid lines correspond to the posterior mean estimations, and the shaded areas correspond to $\lbrace10\%,20\%,...,99\%\rbrace$ credible intervals (around the posterior mean)}.}
    \label{Fig:Entropy1}
\end{figure}

\begin{figure}[H]
    \centering
    \includegraphics[scale=0.68]{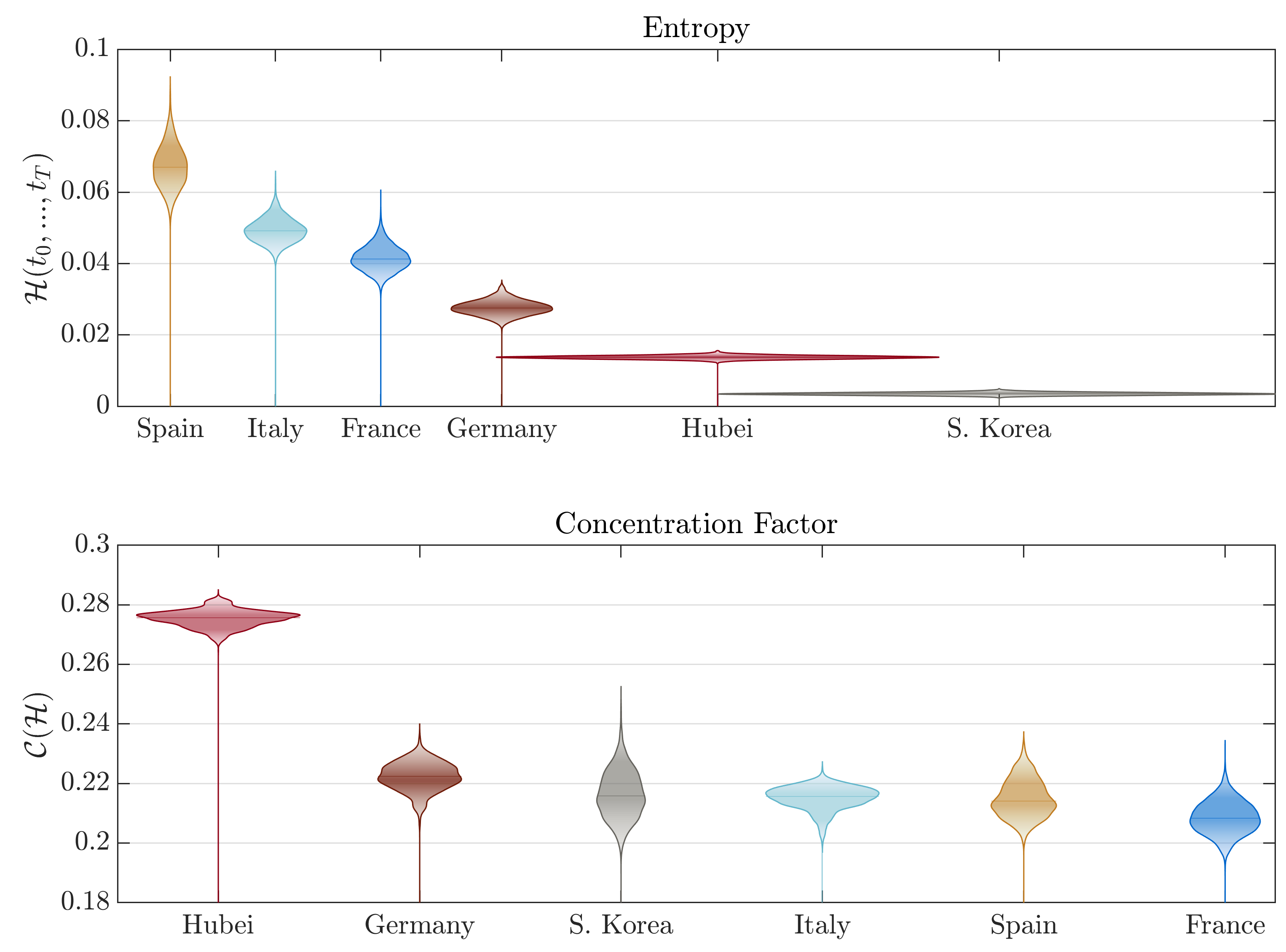}
    \caption{\textbf{The entropies and concentration factors for various regions}. \textit{The figure shows a comparison of total entropies and concentration factors for various regions, with the violins illustrating the posterior distribution}.}
    \label{Fig:Entropy2}
\end{figure}

\noindent After performing model calibrations on datasets of various regions, we present a comparison analysis based on various transmission measures. Figure \ref{Fig:Entropy1} shows the evolution of the entropy rate of COVID-19 outbreaks for each of the regions considered. This graph represents the time evolution of the degree of disorder (in terms of infections and deaths) introduced by the virus in an average statistical individual of the region. This graph reflects features of the daily evolution of infection and recovered/deaths, but it is fundamentally different from the evolution of each compartment. In fact, it has the key property of being objective and comparable between regions. Interestingly, the evolution of the entropy rate has a similar form for each region, but a significant difference in the magnitude of the disorder. In particular, the cumulative integral of the entropy rate represents the change of entropy in the system and, therefore, the total impact in a region. In Figure \ref{Fig:Entropy2}-top panel, we report this impact measure for each of the regions considered. Based on this metric, Spain was the most affected region despite the epidemic wave hit the country later than Italy. On the opposite side, South Korea is the country with the least change in entropy, highlighting an effective combination of policies and cultural habits that limited the impact of the epidemic. This is probably due to the experience gained during the recent 2015 Middle East Respiratory Syndrome coronavirus (MERS-CoV) outbreak \cite{oh2018middle}.  Also, Hubei's reaction, with extreme containment measures, has overall limited the impact of the epidemic. Germany has the smallest total entropy among studied European countries.

Interestingly, the peak of entropy rate for Spain, Italy, and Germany occurred in about the same period but with a different left tail behavior (i.e., in the growing phase). On the other hand, the behavior of the right tail (i.e., the descent phase) is similar, showing a fatter and longer tail. A similar asymmetry can also be observed in Hubei and South Korea. A deviation from this ``classic" behavior is represented by Hubei, which does not show this long tail behavior but has a rather compact and almost symmetric shape.  A surprising result is shown in Figure \ref{Fig:Entropy2}-bottom panel. Although the impact in each country is significantly different, the concentration factor is similar to support the fact that the evolution of COVID-19 is similar for all outbreaks. The Hubei region is slightly deviating from this trend, showing a higher concentration factor corroborating the lack of a right fat tail and, therefore, showing a higher prevalence as an impulse.

\begin{figure}[H]
    \centering
    \includegraphics[scale=0.68]{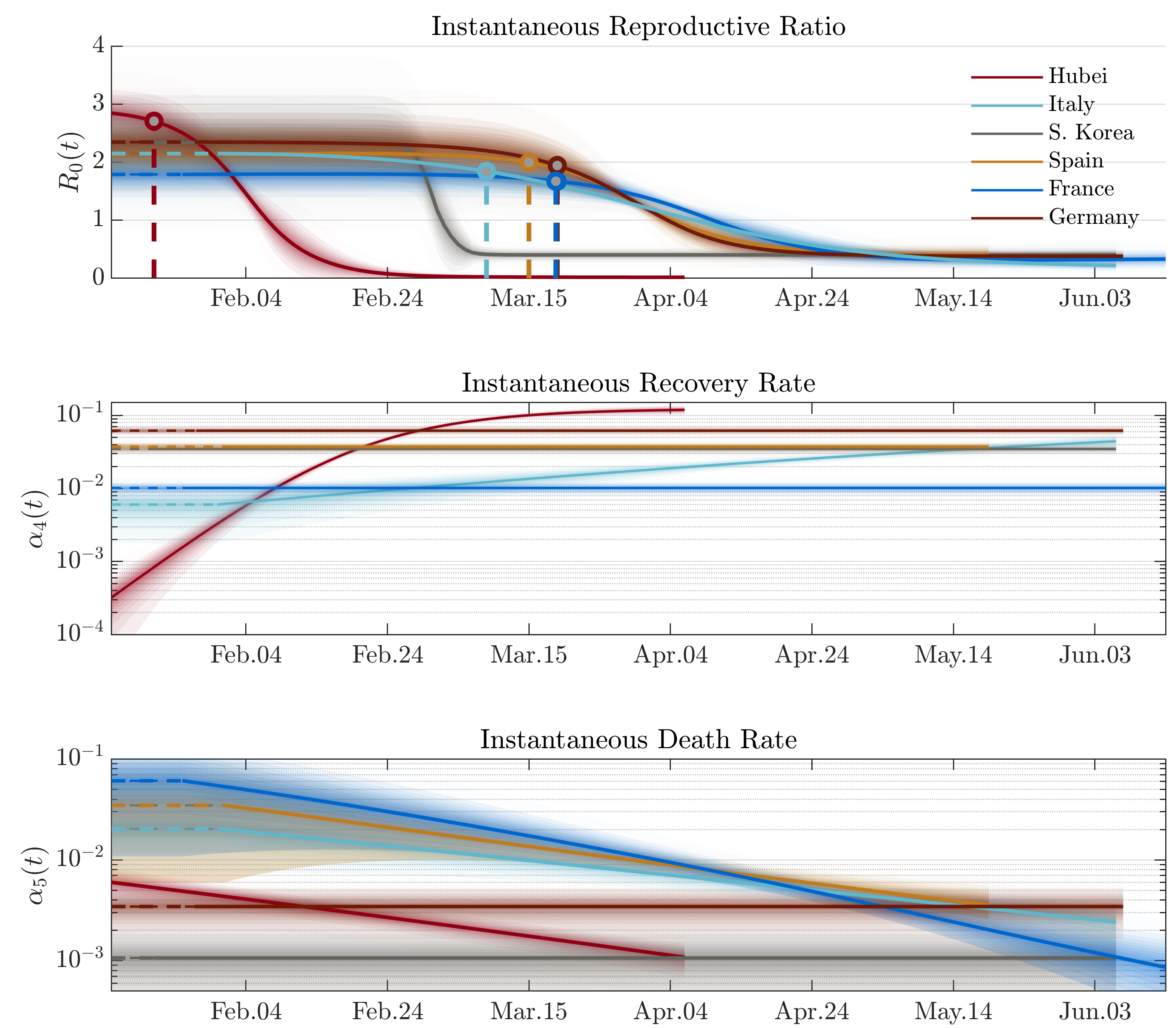}
    \caption{\textbf{The instantaneous reproductive ratio, recovery and death rates for various regions}. \textit{The lockdown date for each region is shown as vertical dashed line. Note that South Korea does not have a lockdown policy}.}
    \label{Fig:R0}
\end{figure}

Figure \ref{Fig:R0} shows a comparison of the instantaneous reproductive ratio, and death rate, together with the date of lockdown in each region. One can infer that the lockdown reduced $R_0(t)$ effectively. However, surprisingly, the most effective decrease has been observed in South Korea were \textit{no} national lockdown has been implemented, but only local containment measures, and massive early stage testing.

\begin{figure}[H]
    \centering
    \includegraphics[scale=0.8]{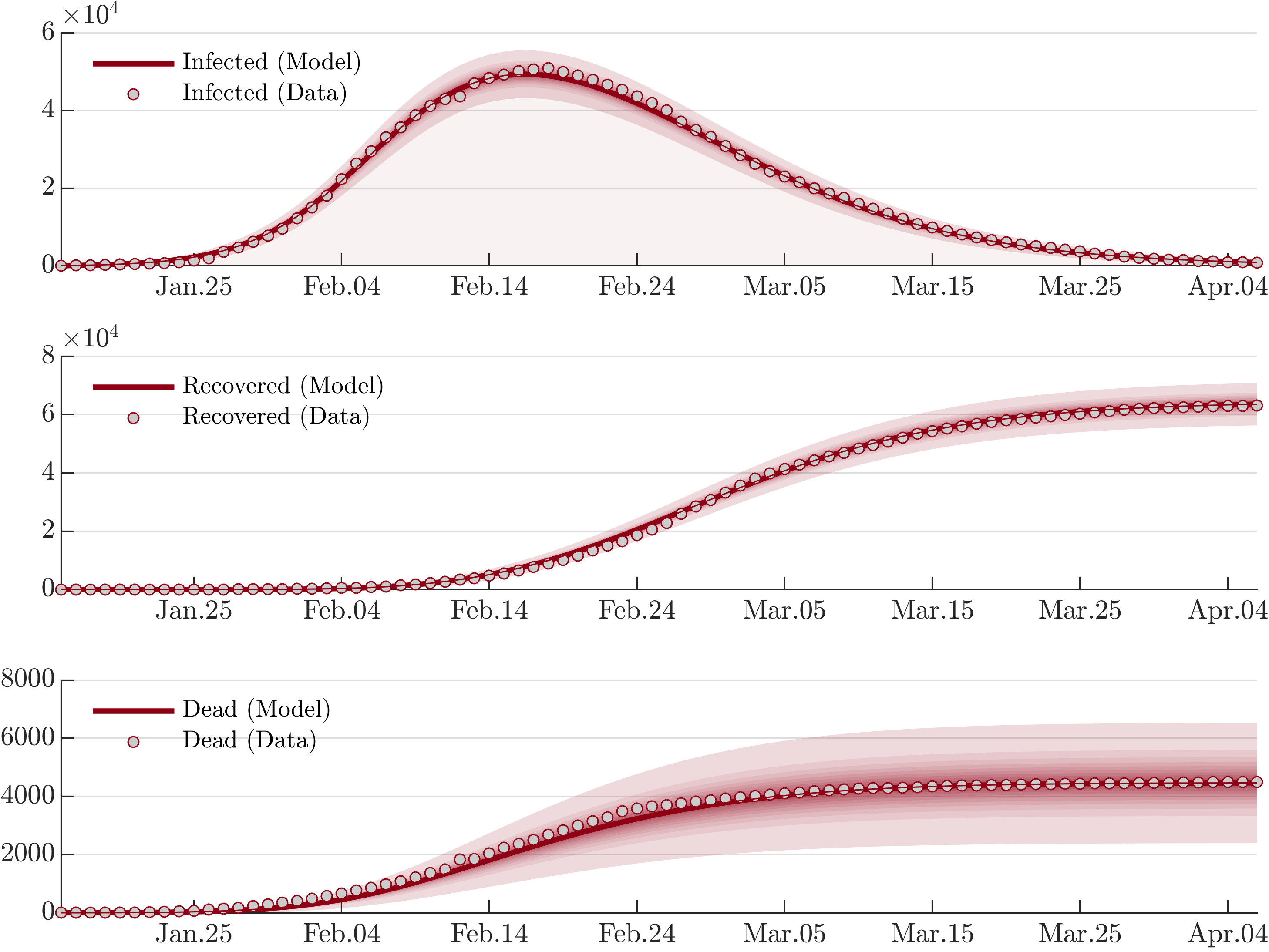}
    \caption{\textbf{Modelling the overall epidemic dynamics of Hubei province with the modified SEIR model}. \textit{The red line corresponds to the posterior mean estimation. The shaded area corresponds to $\lbrace10\%,20\%,...,99\%\rbrace$ credible intervals around the posterior mean. The parameters $\alpha_1(t)$, $\alpha_4(t)$ and $\alpha_5(t)$ are modeled with a single sigmoid basis, and $\alpha_2$ is modeled with a constant variable. The $n_{\epsilon}$ in Eq.\eqref{GaussianKernelLikelihood} is fixed to 100, assuming the error in the increment of infected is of the order of a few hundreds. The figure suggests a highly accurate calibration on data using at most one adaptive basis function for each parameter}.}
    \label{Fig:Hubei}
\end{figure}

It is important to note that the modelling results are associated with the optimized parsimonious model for each region. Specifically, in an optimized parsimonious model the number of time-dependent variables as well as the number of adaptive basis functions for each time-dependent variable are optimized, in the sense that increasing the number would not noticeably improve the calibration accuracy and decreasing the number would significantly degrade the accuracy. Finally, for an illustration on the degree of accuracy the model has achieved, the model calibration results of Hubei is shown in the Figure \ref{Fig:Hubei}. The calibration for the other countries, and their limitations are reported in \ref{Append:ModelResults}.

\subsection{Robustness on the transmission trend}
\noindent A natural concern regarding the discovered transmission trend is that if the trend is a genuine underlying structure of the epidemic, or it is merely some artificial/superficial structures from the specific time-dependent model. It is challenging to (perfectly) resolve this concern because a compartmental model (or any mathematical model) is inevitably an approximation on the real epidemic. Moreover, even an exact model exists, it is still challenging (if not impossible) to accurately identify the model due to the presence of endogenous variables. However, at least we could show that the proposed framework is self-consistent. In \ref{RobustTest}, we simulate artificial epidemics from analytical SEIR laws and investigate if the proposed modelling approach could identify correct transmission trends.


\section{Limitations and future research directions}
\subsection{Incorporating the undetected cases}
\noindent In Section \ref{NumericalTest}, the reported/observed population in each compartment is used to calibrate the model, and the kernel function in Eq.\eqref{GaussianKernelLikelihood} only flattens the likelihood function instead of altering its intrinsic shape. Consequently, the model describes an epidemic scenario consistent with but also confined by the reported cases. An important missing issue to address is to incorporate the undetected cases to fully uncover the magnitude of the epidemic. A practical modeling strategy is to introduce a probability distribution assumption on the (possibly time-dependent) ratio between reported and undetected cases, and rewrite the likelihood function similar to Eq.\eqref{ErrorLikelihood}. Clearly, the critical ingredient is the model assumption on the undetected. The ongoing studies on bloodtest for antibodies of SARS-CoV-2 \cite{NIU} can be useful for this future research direction.
\subsection{Application to more complex compartmental models}
\noindent Depending on the modelling purposes, one could introduce additional compartments, e.g., the tested/suspected, the ICU case, the female and male, the old and young, etc, to study the interactions between different groups. It is also straightforward to include spatial distributed information by including  adjacency and incidence metricises. However, one should be aware that the model variance and the possibility of converging to local insignificant likelihood modes in general would increase with model complexity.  Therefore, it would be crucial to collect robust prior knowledge regarding the modeling parameters.

\section{Conclusions}
\noindent In this study, we have proposed a stochastic compartmental modeling framework of epidemics equipped with entropy-based metrics to measure both the impact and the evolution of a pandemic event. The model belongs to the nonlinear Markov processes class, which allows a robust formulation and a natural setting for developing entropy-based metrics. In addition, we have provided a complete Bayesian inversion scheme to calibrate the model parameters with related uncertainties. Subsequently, we specialized the proposed structure to a modified SEIR model and the COVID-19 pandemic. In particular, we used the framework to investigate six regions: Hubei, South Korea, Italy, Spain, Germany, and France. We showed that the change in entropy in the selected areas (which is associated with the impact of an epidemic) is significantly different. However, it is surprising to note that the dynamic evolution of pandemic waves shows very regular trends and very similar concentration measures.

\section*{Acknowledgement}
\noindent

\bibliography{main}
\appendix
\section{Illustration on the entropy-based measures}\label{Append:Entropy}
\begin{figure}[H]
    \centering
    \includegraphics[scale=0.7]{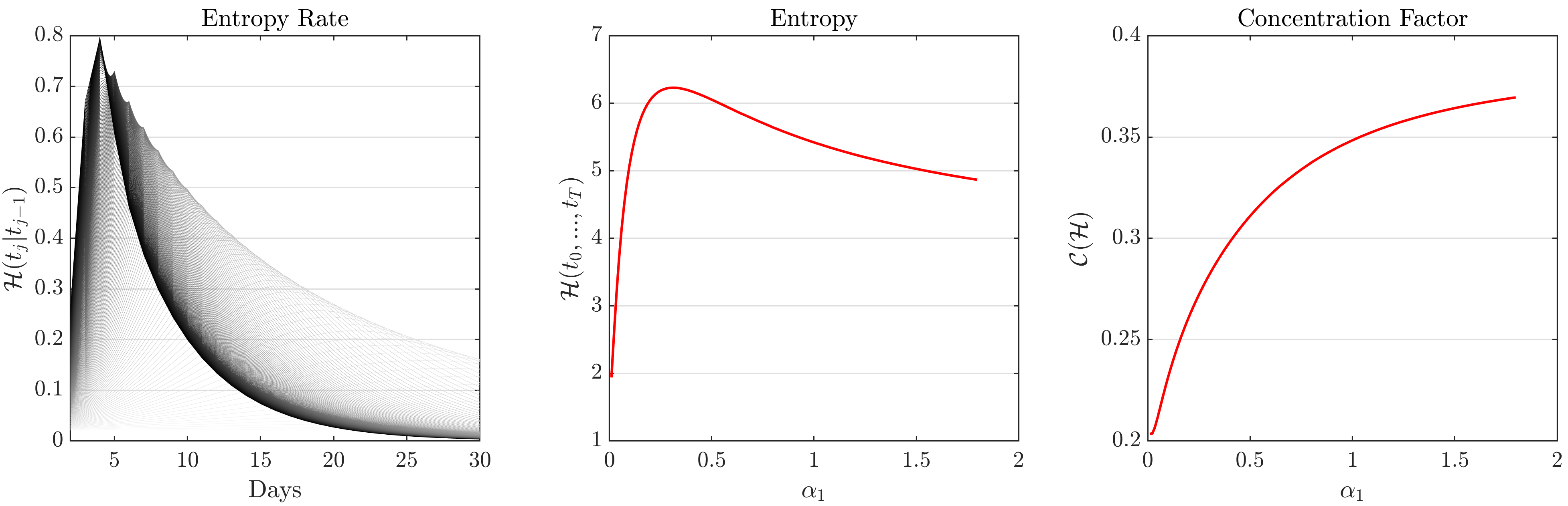}
    \caption{\textbf{Illustration of entropy-based measures with varying $\alpha_1$ values in a stochastic SIR model}. \textit{The left figure illustrates the time evolution of entropy rate with different $\alpha_1$ values, and the lighter to darker curves are associated with lower to higher $\alpha_1$ values. The middle and right figures illustrate the total entropy and concentration factor with respect to $\alpha_1$. It is seen from the figure that as the force of infection grows (while the recovery and death rate are fixed), the evolution of the entropy rate becomes more and more pulse-like. A more important observation is that the entropy may have a peak value due to the trade-off between strong-phase magnitude and  duration of the entropy rate evolution}.}
    \label{Fig:EntropyIllustrate_1}
\end{figure}

\noindent To avoid unnecessary complications, we consider a simple stochastic SIR model with the constant parameters $\bm\alpha=[\alpha_1,\alpha_2,\alpha_3]$ associated with the infection, recovery and death rates, respectively. First, we fix $\alpha_2=0.15$ and $\alpha_3=0.05$, and let $\alpha_1$ vary within $[0.2,2]$, so that the basic reproductive ratio varies within $[1,10]$. The initial condition is set to $\bm P(t_0)=[0.99,0.01,0,0]$. Figure \ref{Fig:EntropyIllustrate_1} illustrates the entropy rate, total entropy and concentration factor.

\begin{figure}[H]
    \centering
    \includegraphics[scale=0.7]{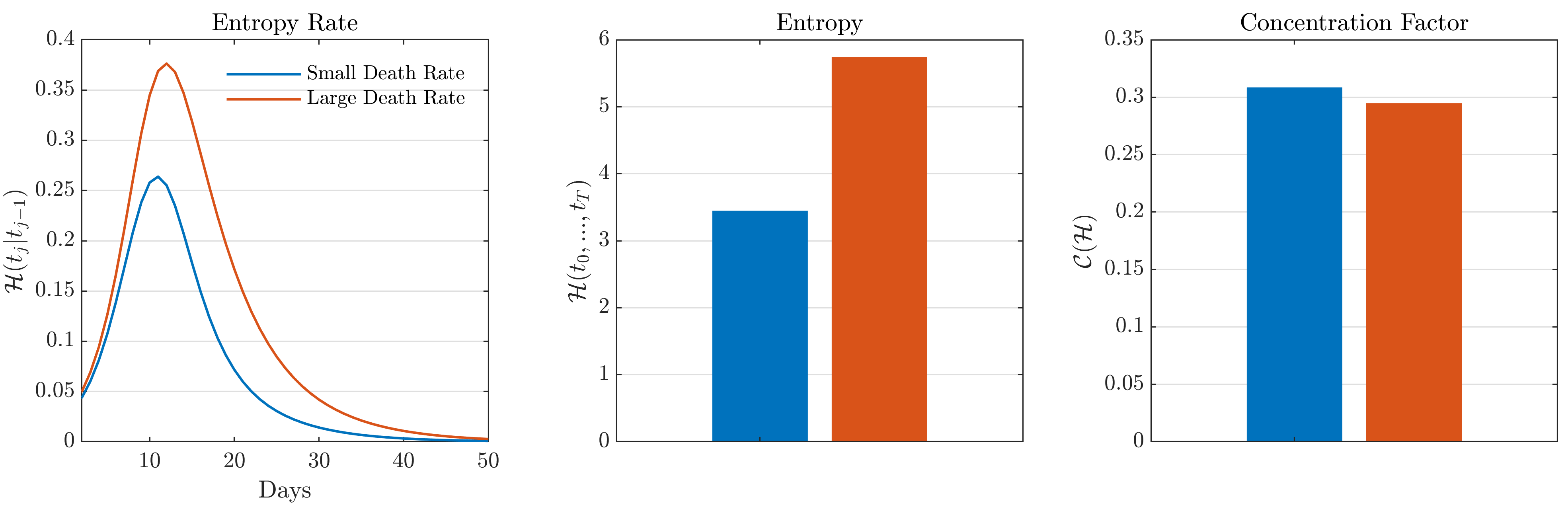}
    \caption{\textbf{Illustration of the entropy-based measures using the reshaped $\bm P$ and $\bm S$}. \textit{The figure suggests that when the infected and recovered states are merged into a compound state, the contribution (to the entropy rate and entropy) from the death cases is increased. The concentration factor is not sensitive to this technique. Note that without using a reshaped $\bm P$ and $\bm S$, the entropy-based measures for the two systems will be exactly the same}.}
    \label{Fig:EntropyIllustrate_2}
\end{figure}

Figure \ref{Fig:EntropyIllustrate_1} also implies that the basic reproductive ratio and the entropy-based measures describe different aspect of the epidemic dynamics, and they do not have an one-on-one mapping. Note that in Figure \ref{Fig:EntropyIllustrate_1} the entropy-based measures are computed using the original distribution vector $\bm P$ and stochastic matrix $\bm S$. In Section \ref{Sec:Entropy} and Section \ref{Sec:Transmission} it is mentioned that $\bm P$ and $\bm S$ can be reshaped to highlight the contribution from the death cases. Figure \ref{Fig:EntropyIllustrate_2} illustrates the effect of this technique. Specifically, two SIR models with the same initial conditions $\bm P(t_0)=[0.99,0.01,0,0]$ are considered. The first model has $\bm\alpha=[0.6,0.19,0.01]$, and the second model is obtained by swapping the recovery and death rates of the first model. The basic reproductive ratios for the two models are both $3.0$.

\section{Results of model calibration}\label{Append:ModelResults}
\begin{figure}[H]
    \centering
    \includegraphics[scale=0.8]{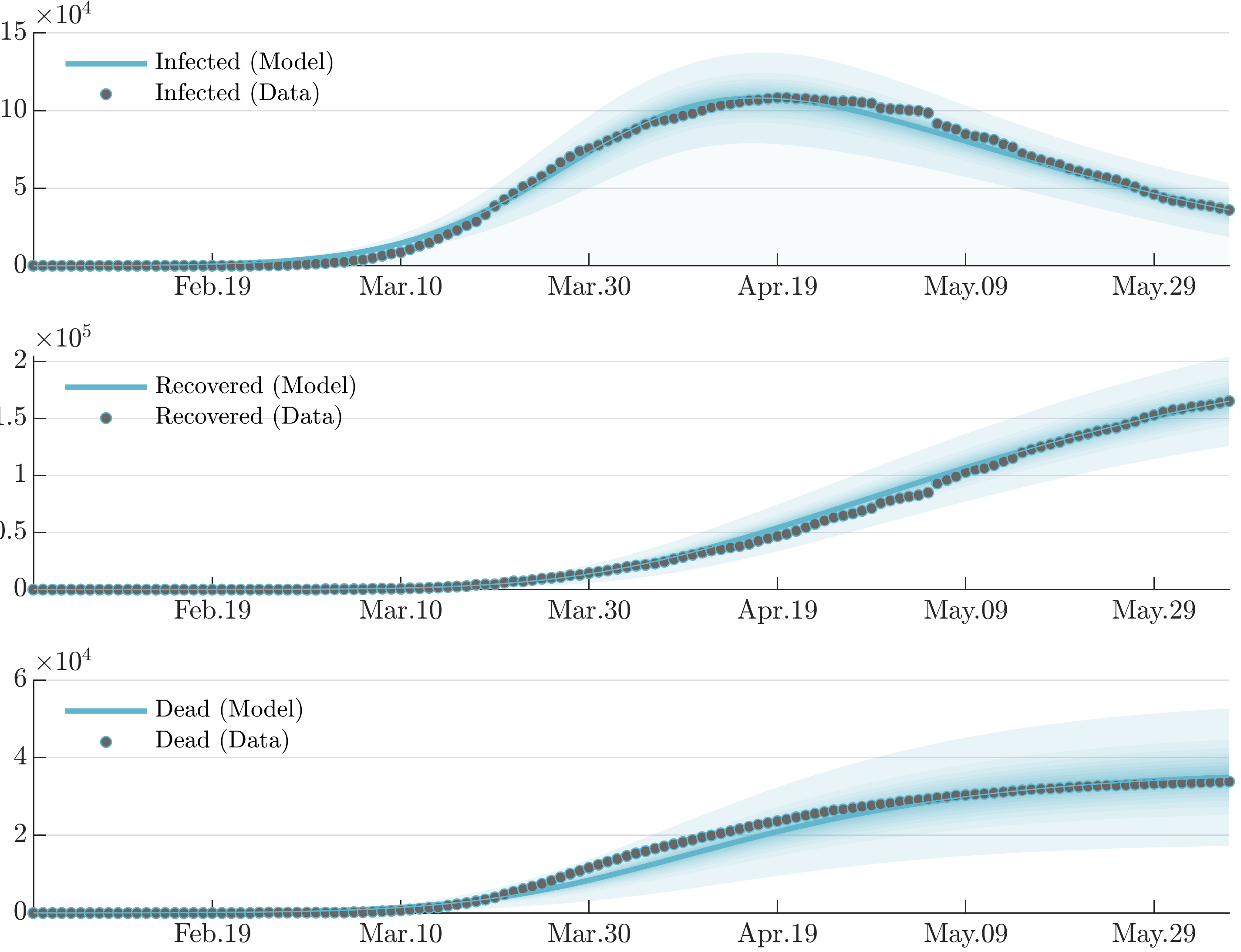}
    \caption{\textbf{Modelling the overall epidemic dynamics of Italy with the modified SEIR model}. \textit{The $\alpha_1(t)$, $\alpha_4(t)$ and $\alpha_5(t)$ are modelled with a single sigmoid basis, and $\alpha_2(t)$ is modelled as a constant variable. The $n_{\epsilon}$ in Eq.\eqref{GaussianKernelLikelihood} is fixed to 1000, assuming the error in the increment of infected is in the order of a few thousands}.}
    \label{Fig:Italy}
\end{figure}

\begin{figure}[H]
    \centering
    \includegraphics[scale=0.8]{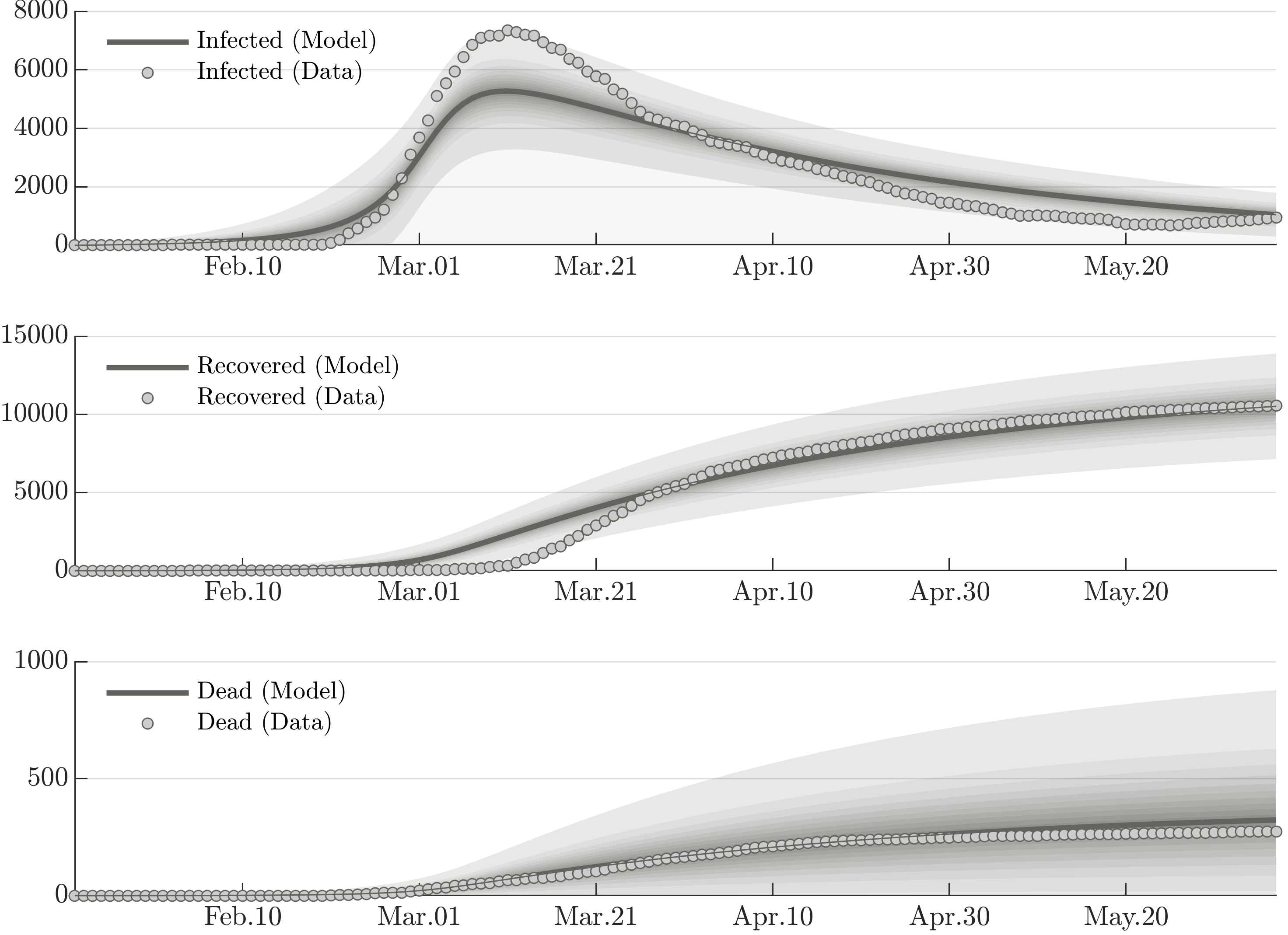}
    \caption{\textbf{Modelling the overall epidemic dynamics of South Korea with the modified SEIR model}. \textit{The $\alpha_1(t)$ is modelled with a single sigmoid basis, and $\alpha_2(t)$, $\alpha_4(t)$ and $\alpha_5(t)$ are modelled as constant variables. The $n_{\epsilon}$ in Eq.\eqref{GaussianKernelLikelihood} is fixed to 100, assuming the error in the increment of infected is of the order of a few hundreds}.}
    \label{Fig:Korea}
\end{figure}

\begin{figure}[H]
    \centering
    \includegraphics[scale=0.8]{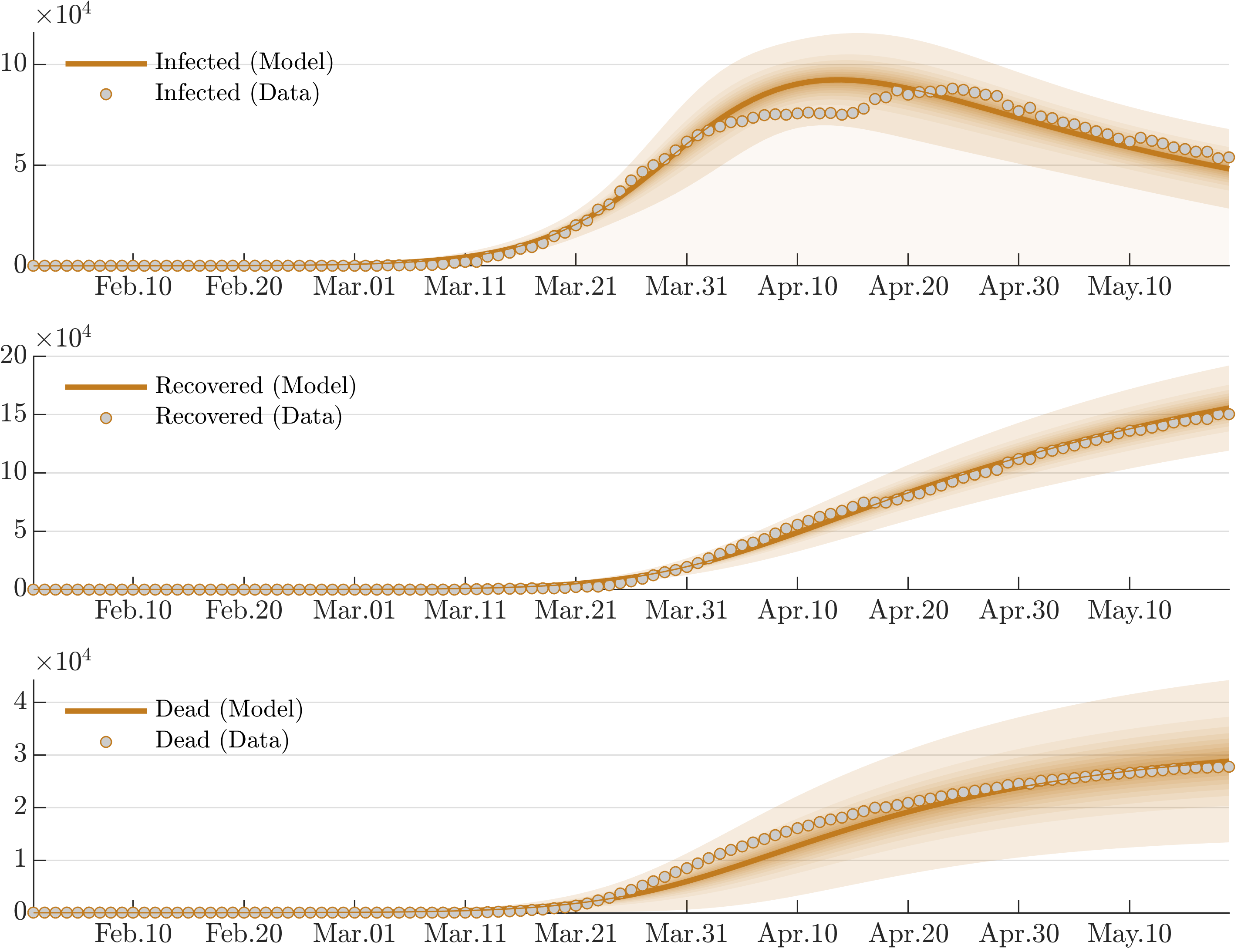}
    \caption{\textbf{Modelling the overall epidemic dynamics of Spain with the modified SEIR model}. \textit{The $\alpha_1(t)$ and $\alpha_5(t)$ and are modelled with a single sigmoid basis, and $\alpha_2$ and $\alpha_4$ are modelled as constant variables. The $n_{\epsilon}$ in Eq.\eqref{GaussianKernelLikelihood} is fixed to 1000, assuming the error in the increment of infected is in the order of a few thousands}.}
    \label{Fig:Spain}
\end{figure}

\begin{figure}[H]
    \centering
    \includegraphics[scale=0.8]{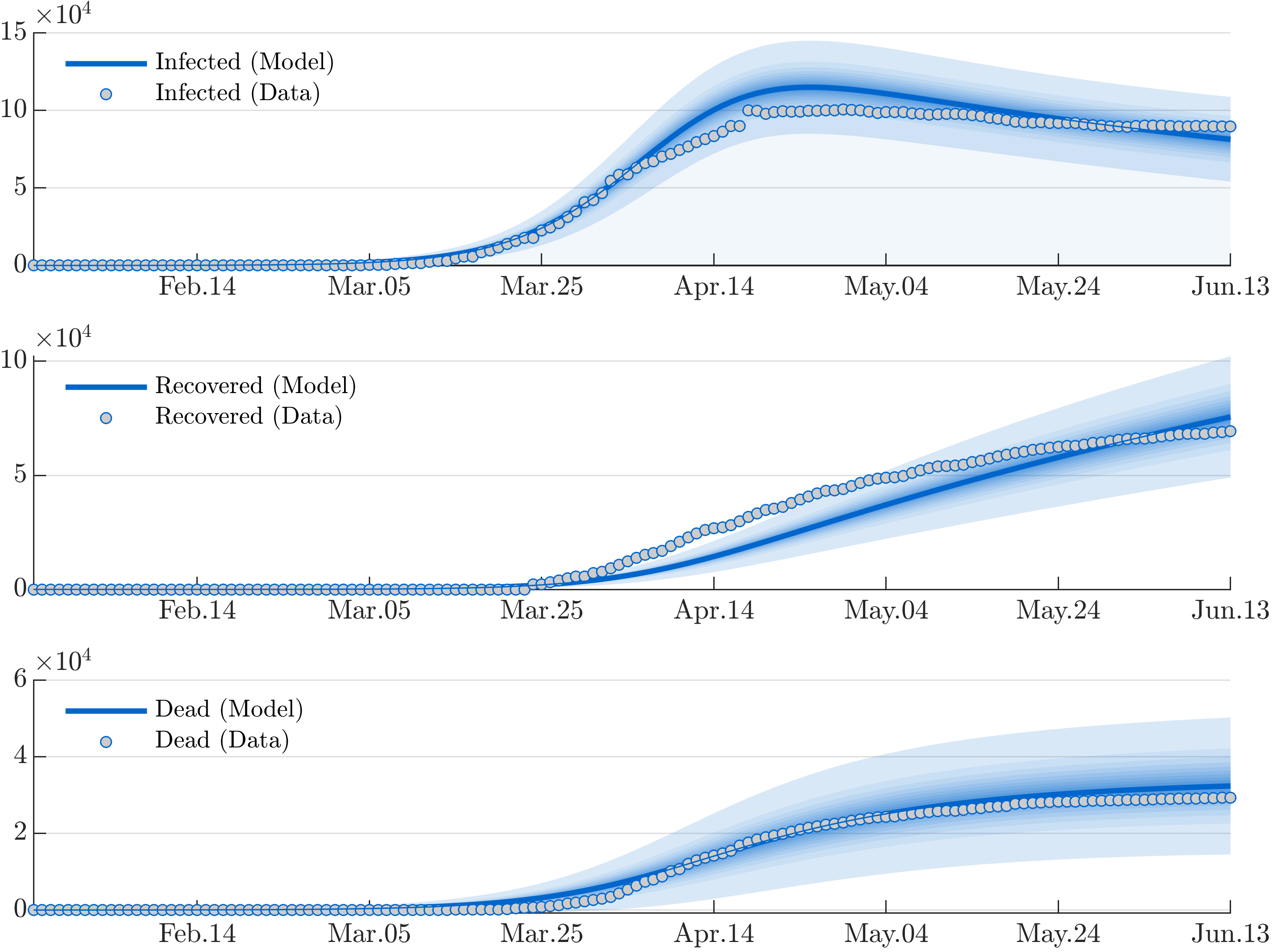}
    \caption{\textbf{Modelling the overall epidemic dynamics of France with the modified SEIR model}. \textit{The $\alpha_1(t)$ and $\alpha_5(t)$ and are modelled with a single sigmoid basis, and $\alpha_2$ and $\alpha_4$ are modelled as constant variables. The $n_{\epsilon}$ in Eq.\eqref{GaussianKernelLikelihood} is fixed to 1000, assuming the error in the increment of infected is of the order of a few thousands}.}
    \label{Fig:France}
\end{figure}

\begin{figure}[H]
    \centering
    \includegraphics[scale=0.8]{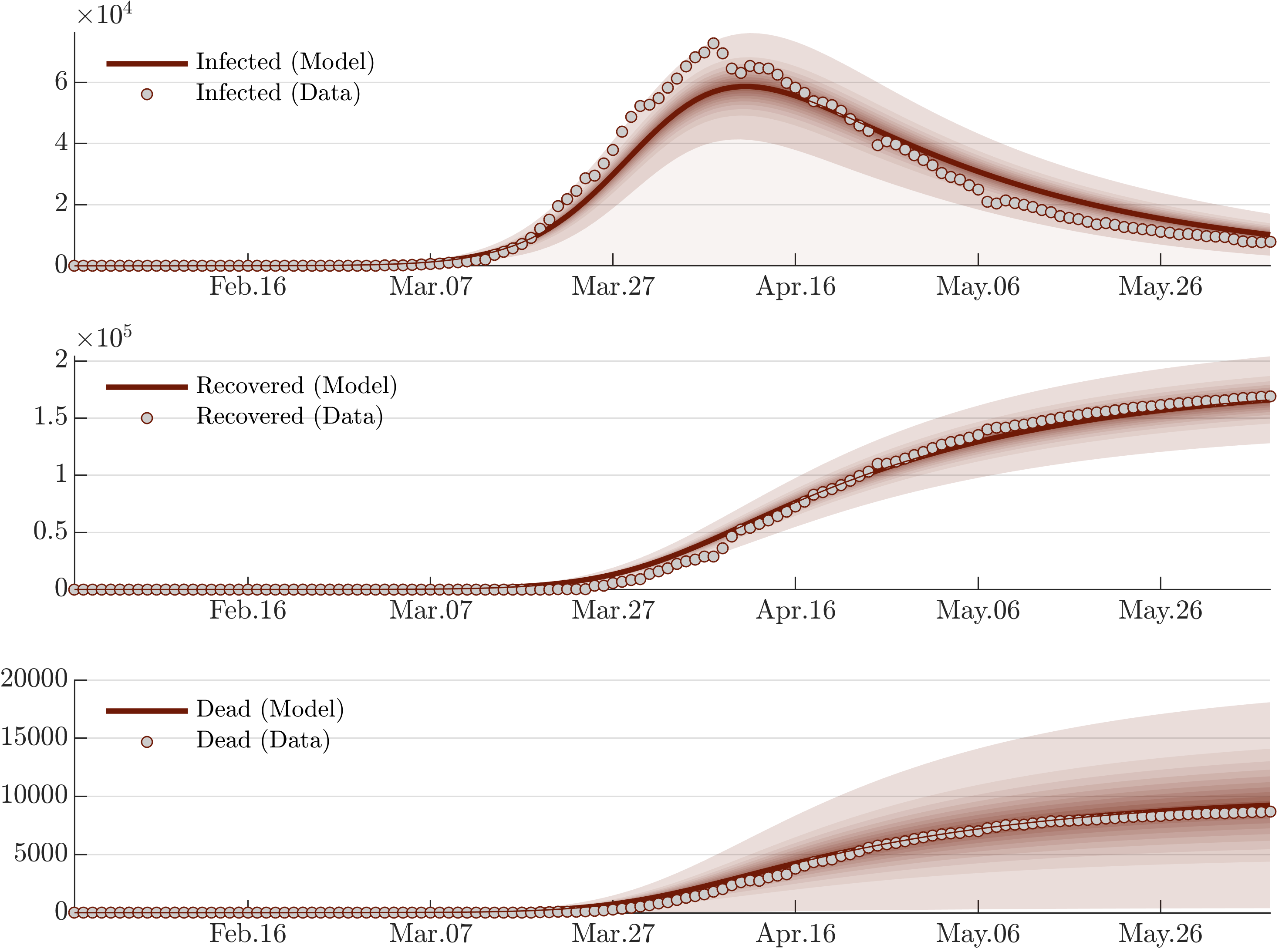}
    \caption{\textbf{Modelling the overall epidemic dynamics of Germany with the modified SEIR model}. \textit{The $\alpha_1(t)$ is modelled with a single sigmoid basis, and $\alpha_2(t)$, $\alpha_4(t)$ and $\alpha_5(t)$ are modelled as constant variables. The $n_{\epsilon}$ in Eq.\eqref{GaussianKernelLikelihood} is fixed to 1000, assuming the error in the increment of infected is of the order of a few thousands}.}
    \label{Fig:Germany}
\end{figure}

\noindent The model calibration results for Italy, South Korea, Spain, France and Germany are shown in Figure \ref{Fig:Italy}-Figure \ref{Fig:Germany}. In the figures, the solid lines correspond to posterior mean estimations, and the shaded areas correspond to $\lbrace10\%,20\%,...,99\%\rbrace$ credible intervals around the posterior mean.

\section{Results on robustness/self-consistent test on the transmission trend}\label{RobustTest}
\begin{figure}[H]
    \centering
    \includegraphics[scale=0.8]{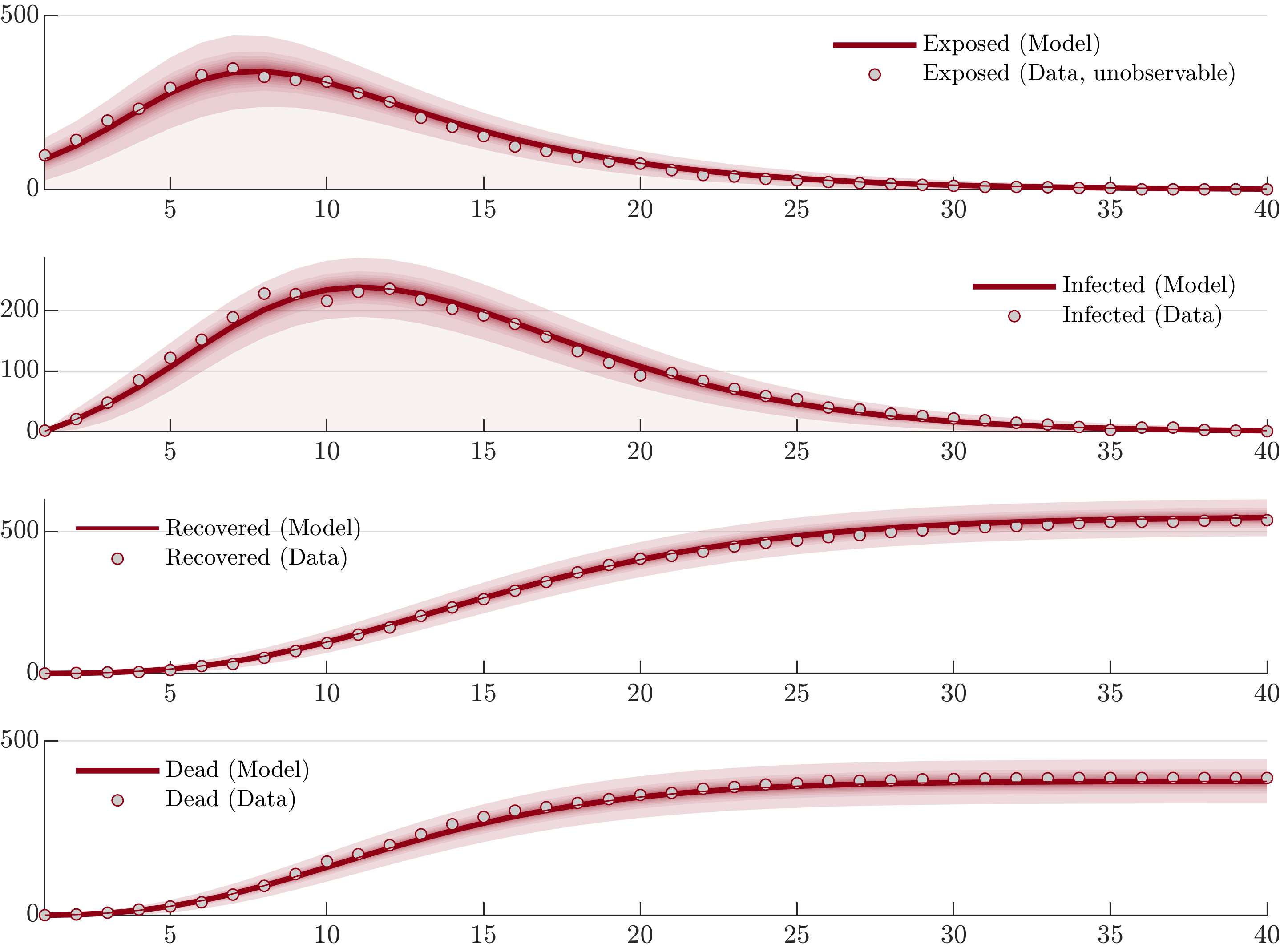}
    \caption{\textbf{The calibrated time-dependent modified SEIR model}. \textit{The $\alpha_1(t)$, $\alpha_2(t)$, $\alpha_4(t)$ and $\alpha_5(t)$ are modelled with a single sigmoid basis, and $\alpha_3$ is modelled as a constant variable. The $n_{\epsilon}$ in Eq.\eqref{GaussianKernelLikelihood} is fixed to 1, assuming no error in the dataset. The figure suggests an accurate calibration. Even the unobserved exposed population and its initial condition are accurately identified. However, this is because the dataset is generated from an analytical model. For real dataset where a mathematical model is only an approximation, it is preferable to use real clinical data to specify the mean incubation period and its epistemic uncertainty (as it is performed in the main text)}.}
    \label{Fig:TestSEIR1}
\end{figure}

\begin{figure}[H]
    \centering
    \includegraphics[scale=0.8]{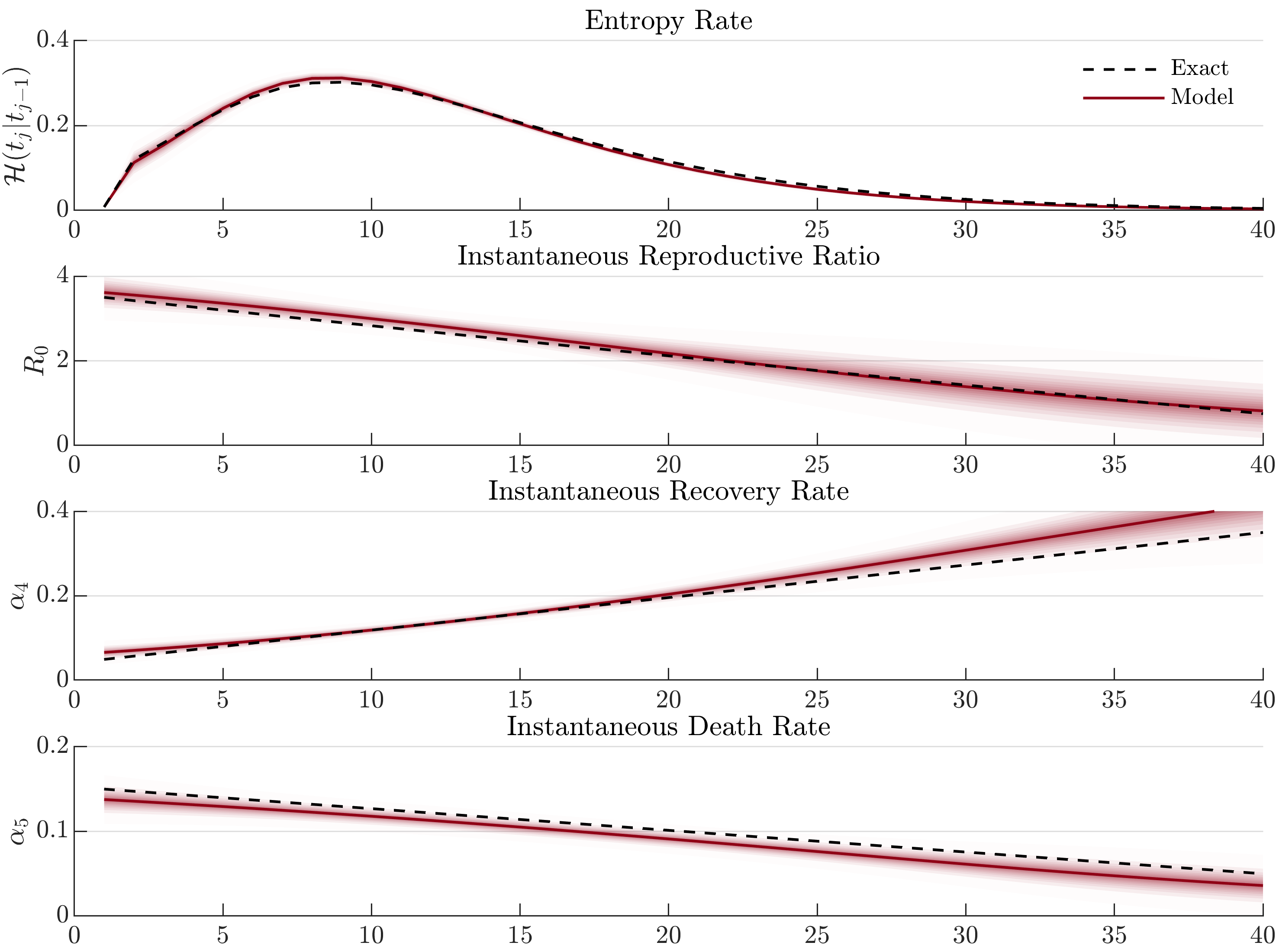}
    \caption{\textbf{The transmission properties of the time-dependent SEIR model compared with the exact values}. \textit{The figure suggests that the transmission trends (in terms of the posterior mean estimation) are in general identified accurately. The moderate bias is mainly caused by the inherent variability of the stochastic simulation (see the following verification)}.}
    \label{Fig:TestSEIR2}
\end{figure}

\noindent We simulate a random epidemic scenario from a modified stochastic SEIR model with parameters $\bm\alpha=[\alpha_1(t),0.10,0.20,\alpha_4(t),\alpha_5(t)]$. The duration of the simulation is set to 40 days. The time-dependent parameters are specified as $\alpha_1(t)=0.6-0.5t/40$, $\alpha_4(t)=0.05+0.30t/40$, and $\alpha_5(t)=0.15-0.10t/40$, and consequently the instantaneous reproductive ratio decreases from $3.50$ to $0.75$ as $t$ grows from 0 to 40 (days). The population size is fixed to $N=10^3$, and the initial condition is set to $\bm P(t_0)=[1-101/N,100/N,1/N,0,0]$. The specification of a relatively large initial condition for the unobservable exposed state is to ``challenge'' the proposed approach and investigate if the exposed population can be accurately identified even without observing it. We calibrate a time-dependent modified SEIR model and the results are illustrated in Figure \ref{Fig:TestSEIR1} and Figure \ref{Fig:TestSEIR2}.

\begin{figure}[H]
    \centering
    \includegraphics[scale=0.8]{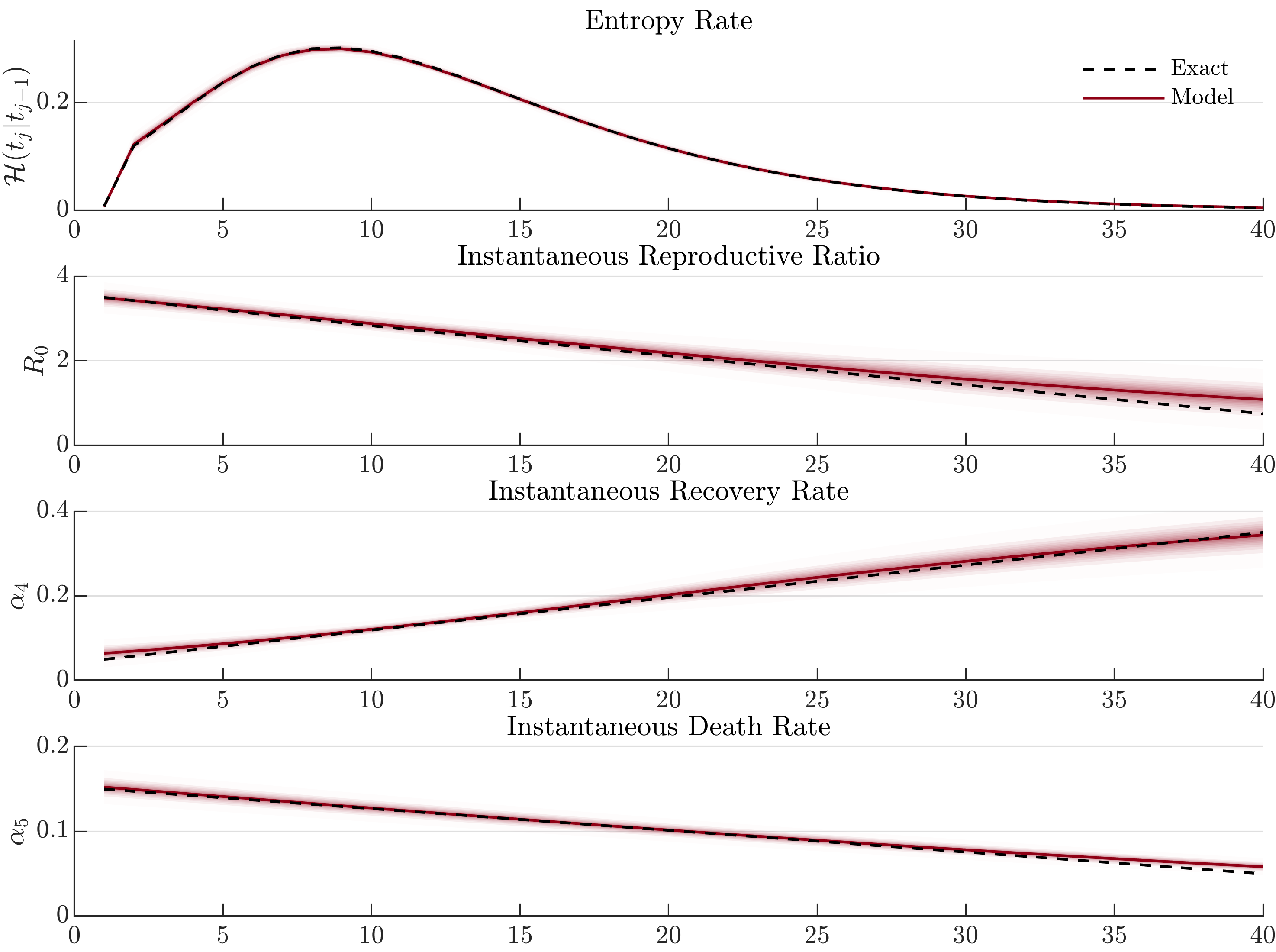}
    \caption{\textbf{The transmission properties of the time-dependent SEIR model calibrated on the expectation of the analytical model}. \textit{Compared with Figure \ref{Fig:TestSEIR2}, it is seen that the bias is noticeably decreased}.}
    \label{Fig:TestSEIR3}
\end{figure}

Finally, to verify that the bias in Figure \ref{Fig:TestSEIR2} is mainly caused by the inherent stochastic variation of the model, we calibrate the model on the expectation of the analytical model, and the results of transmission properties are illustrated in Figure \ref{Fig:TestSEIR3}.

\end{document}